# Cardiac Digital Twin Pipeline for Virtual Therapy Evaluation

Julia Camps[1][*][a], Zhinuo Jenny Wang[1][*][a], Ruben Doste[1], Maxx Holmes[1], Brodie Lawson[2], Jakub Tomek[1], Kevin Burrage[2], Alfonso Bueno-Orovio[1], Blanca Rodriguez[1]

[1]University of Oxford, Oxford, United Kingdom.

[2]Queensland University of Technology, Brisbane, Australia.

**\* Corresponding authors:**

Julia Camps                           Zhinuo Jenny Wang
julcamps@gmail.com          jenny.wang@cs.ox.ac.uk

[a] Julia Camps and Zhinuo Jenny Wang made equal contributions to this work.

Cardiac digital twin; Precision cardiology; Virtual therapy evaluation; reaction-Eikonal model; Monodomain model; Cardiac Magnetic Resonance; Electrocardiogram

## Abstract

Cardiac digital twins are computational tools capturing key functional and anatomical characteristics of patient hearts for investigating disease phenotypes and predicting responses to therapy. When paired with large-scale computational resources and large clinical datasets, digital twin technology can enable virtual clinical trials on virtual cohorts to fast-track therapy development. Here, we present an automated pipeline for personalising ventricular anatomy and electrophysiological function based on routinely acquired cardiac magnetic resonance (CMR) imaging data and the standard 12-lead electrocardiogram (ECG).

Using CMR-based anatomical models, a sequential Monte-Carlo approximate Bayesian computational inference method is extended to infer electrical activation and repolarisation characteristics from the ECG. Fast simulations are conducted with a reaction-Eikonal model, including the Purkinje network and biophysically-detailed subcellular ionic current dynamics for repolarisation. For each patient, parameter uncertainty is represented by inferring a population of ventricular models rather than a single one, which means that parameter uncertainty can be propagated to therapy evaluation. Furthermore, we have developed techniques for translating from reaction-Eikonal to monodomain simulations, which allows more realistic simulations of cardiac electrophysiology. The pipeline is demonstrated in a healthy female subject, where our inferred reaction-Eikonal models reproduced the patient's ECG with a Pearson's correlation coefficient of 0.93, and the translated monodomain simulations have a correlation coefficient of 0.89. We then apply the effect of Dofetilide to the monodomain population of models for this subject and show dose-dependent QT and T-peak to T-end prolongations that are in keeping with large population drug response data.

The methodologies for cardiac digital twinning presented here are a step towards personalised virtual therapy testing and can be easily scaled to generate virtual populations for clinical trials to fast-track drug safety evaluation. The tools developed for this paper are open-source, documented, and made publicly available.

**Graphical abstract**

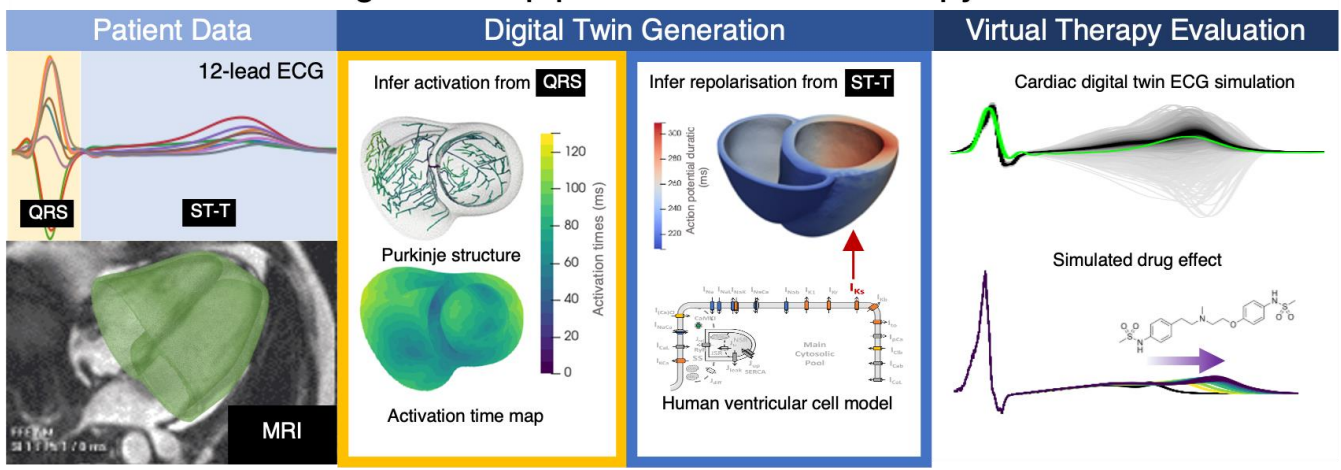

**Highlights**

- Novel pipeline for generating cohorts of cardiac digital twins with personalised activation and repolarisation characteristics from clinical CMR and ECG data.
- Human-based reaction-Eikonal implementation to enable fast digital twinning.
- Effective strategy for translation from reaction-Eikonal to monodomain simulations for drug evaluation.
- Parameter uncertainty propagation for virtual therapy testing.
- Compilation of experimental and clinical evidence on ventricular APD and repolarisation gradients.
- An open-source implementation of a cardiac digital twinning pipeline is available.

# 1 Introduction

Significant differences in cardiac anatomy and electrophysiological function in the human population drive the need for a precision medicine approach when developing cardiac therapies (Antman & Loscalzo, 2016). Cardiac digital twin is an emerging paradigm describing a suite of tools that continuously and coherently integrates patient data to produce virtual hearts that evolve with their 'twin' to help realise the vision of precision medicine in cardiology (Corral-Acero et al., 2020; Gillette et al., 2021). For such cardiac digital twin technologies to be useful for therapy development, computational modelling choices should ensure that the key therapy targets are mathematically represented in sufficient detail so that 1) virtual predictions of therapy outcomes are mechanistic, relevant, and trustworthy, and 2) any parameter uncertainties due to varying levels of identifiability in the digital twin are considered. In addition, with the advent of exascale computing (Hoekstra et al., 2019) and Big Data in healthcare (Littlejohns et al., 2020), cardiac digital twin technology should be easily scalable to enable virtual cohorts for *in silico* clinical trials (Passini et al., 2017; Musuamba et al., 2021; Dasi et al., 2022; Roney et al., 2022; Fassina et al., 2023). Here, we present a cardiac digital twin generation pipeline developed with these goals in mind, using routinely acquired cardiac magnetic resonance (CMR) data and 12-lead electrocardiograms (ECG).

Drug electrophysiological safety and efficacy evaluation is based on its effect on cardiac activation and repolarisation properties. The ECG encodes information about these properties: the QRS complex of the ECG reflects the activation pattern, while the ST-T ECG segment comprises information on spatial heterogeneities in repolarisation and action potential duration (APD) (Szentadrassy et al., 2005; Opthof et al., 2017)Repolarisation heterogeneities are underpinned by a complex interplay of subcellular ionic current dynamics (Szentadrassy et al., 2005; Opthof et al., 2017), which are altered by antiarrhythmic drugs, such as Dofetilide. Therefore, phenomenological models that do not explicitly describe these ionic currents are of limited relevance in the context of virtual drug evaluations (Gillette et al., 2021, 2022). However, the high computational cost of simulating electrotonic coupling with human-based ionic current dynamics using either the gold standard reaction-diffusion equations (Potse et al., 2006) or reaction-diffusion-Eikonal models (Neic et al., 2017) limits the scalability of the digital twin generation pipeline. Therefore, this paper presents a strategy that leverages the benefits of both phenomenological and biophysically detailed models to enable scalable and relevant cardiac digital twins for therapy evaluation.

## 2   Methods

As an overview, this study extends our previous pipeline (Banerjee et al., 2021; Camps et al., 2021, 2022, 2023) to automatically infer the activation and repolarisation properties of a biophysically detailed cardiac electrophysiological model from CMR and 12-lead ECG (Figure 1). Virtual drug evaluation is demonstrated by simulating the dose-dependent effects of Dofetilide application, and the simulated results were compared with clinical data (Johannesen et al., 2014; Vicente et al., 2015). The pipeline is demonstrated in a healthy female subject whose biventricular cardiac geometry was extracted from CMR (Mincholé et al., 2019; Banerjee et al., 2021). Two tetrahedral meshes were generated for their ventricular geometry with resolutions of 1500 and 900 microns for reaction-Eikonal and monodomain simulations, respectively.

Fast inference was enabled by developing a reaction-Eikonal model that could be easily translated to monodomain simulations through mimicking electrotonic coupling effects. The inference of the activation properties (that includes earliest activation sites and conduction velocities) was performed first using only QRS complex data (Camps et al., 2021, 2022, 2023). Smooth spatial variation (heterogeneity) in APDs was then independently inferred from the ST-T ECG signals.

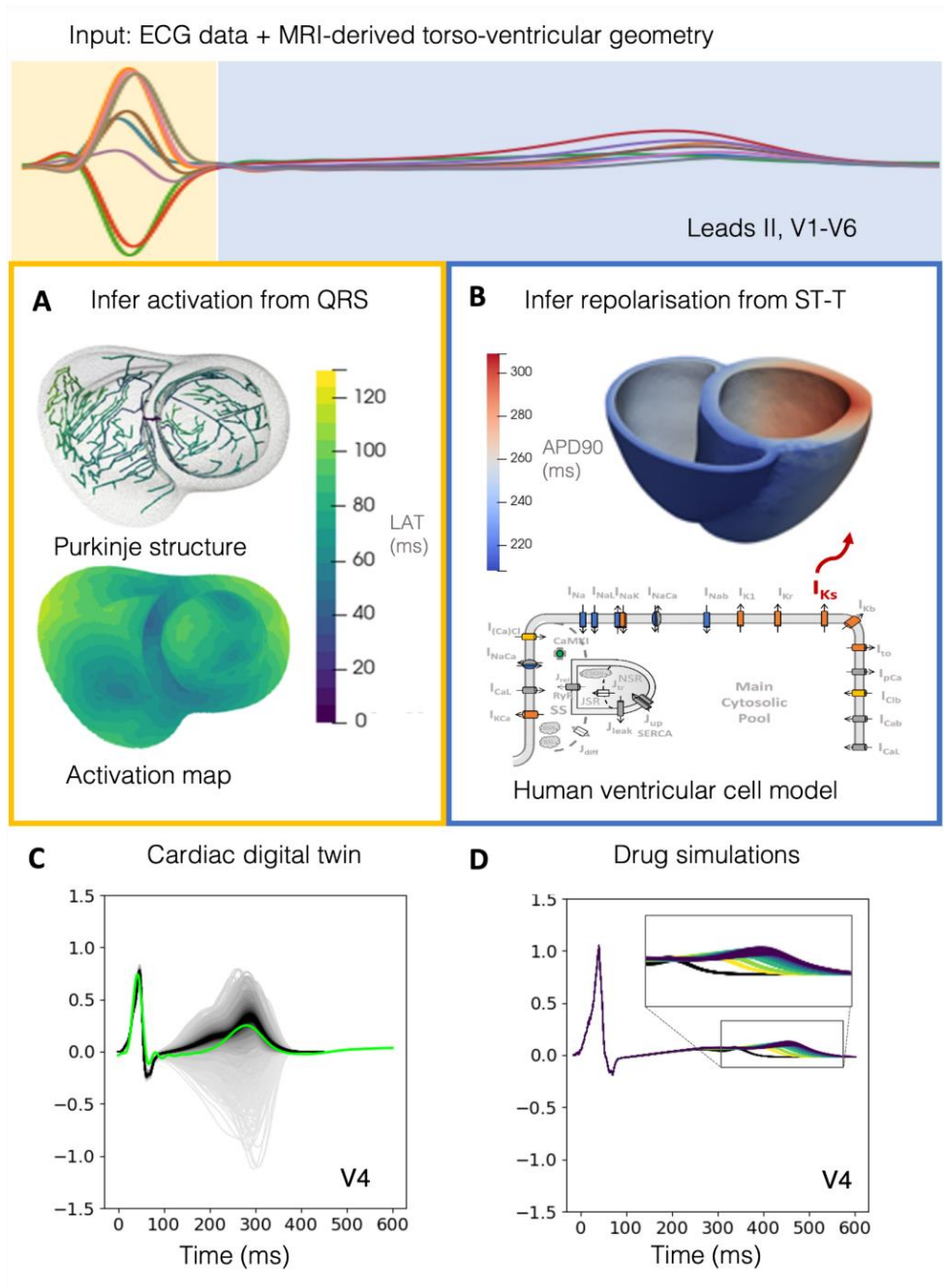

Figure 1: Overview of our cardiac digital twin personalisation pipeline. (A) Firstly, the pipeline infers the conduction speeds and the Purkinje-informed locations of earliest activation on the endocardial surface of the heart from matching QRS simulations to the subject's ECG. (B) The pipeline then infers the spatial heterogeneity of the slow

delayed rectifier potassium current ($I_{Ks}$), which underpins repolarisation heterogeneity, by matching simulations of the ST-T segment to the subject's ECG. (C) This process produces the cardiac digital twin, which contains the final inferred population of models (black traces) that matches the subject's ECG (green trace). (D) These models are then translated to monodomain simulations, and Dofetilide application is added to evaluate the ability of the model to accurately predict drug effects, where the black traces are the baseline and the coloured traces are at varying doses of Dofetilide application.

## 2.1    Data preparation and representation for the digital twin pipeline

The pipeline takes as input the subject-specific female CMR-derived biventricular geometry (Mincholé et al., 2019; Banerjee et al., 2021), the subject's 12-lead ECG recordings, and a rule-based description of the fibre, sheet, and sheet-normal vector fields (Streeter et al., 1969; Levrero-Florencio et al., 2020).

## 2.2    Action potential duration gradients model

Spatial heterogeneity in APD, defined at 90% of repolarisation (APD90 or just APD for this study), was represented by a weighted linear sum of four ventricular coordinates (Gillette et al., 2021) calculated using the position of a mesh node in three-dimensional space, $\boldsymbol{x}$. These coordinates are the apex-to-base coordinate $ab(\boldsymbol{x})$ as defined by Cobiveco (Schuler et al., 2021), the transmural coordinate $tm(\boldsymbol{x})$ as in Bayer et al. (2018), and the transventricular coordinate (left-to-right ventricle) $tv(\boldsymbol{x})$, and the posterior-to-anterior $pa(\boldsymbol{x})$ coordinate. The transventricular and posterior-to-anterior coordinates were obtained by normalising the projections of the nodal coordinates along the transventricular and posterior-to-anterior directions, which were defined as the normal to the septal surface (transventricular) and the cross product of the septal surface and basal surfaces normals (posterior-to-anterior). These normal directions were obtained as in (Doste et al., 2019). These coordinates are used to define a spatially varying APD field (Figure 2) that ranges between a specified minimum ($APD_{min}$) and maximum ($APD_{max}$) value, as follows

$$APD(\boldsymbol{x}) = \left(\frac{q(\boldsymbol{x}) - q(\boldsymbol{x})_{min}}{q(\boldsymbol{x})_{max} - q(\boldsymbol{x})_{min}}\right)(APD_{max} - APD_{min}) + APD_{min}$$

$$q(\boldsymbol{x}) = g_{ab}ab(\boldsymbol{x}) + g_{tm}tm(\boldsymbol{x}) + g_{pa}pa(\boldsymbol{x}) + g_{tv}tv(\boldsymbol{x})$$

(1)

The weighting parameters $g_{ab}$, $g_{tm}$, $g_{pa}$, and $g_{tv}$ control the relative magnitude of the APD gradient in their respective coordinate directions.

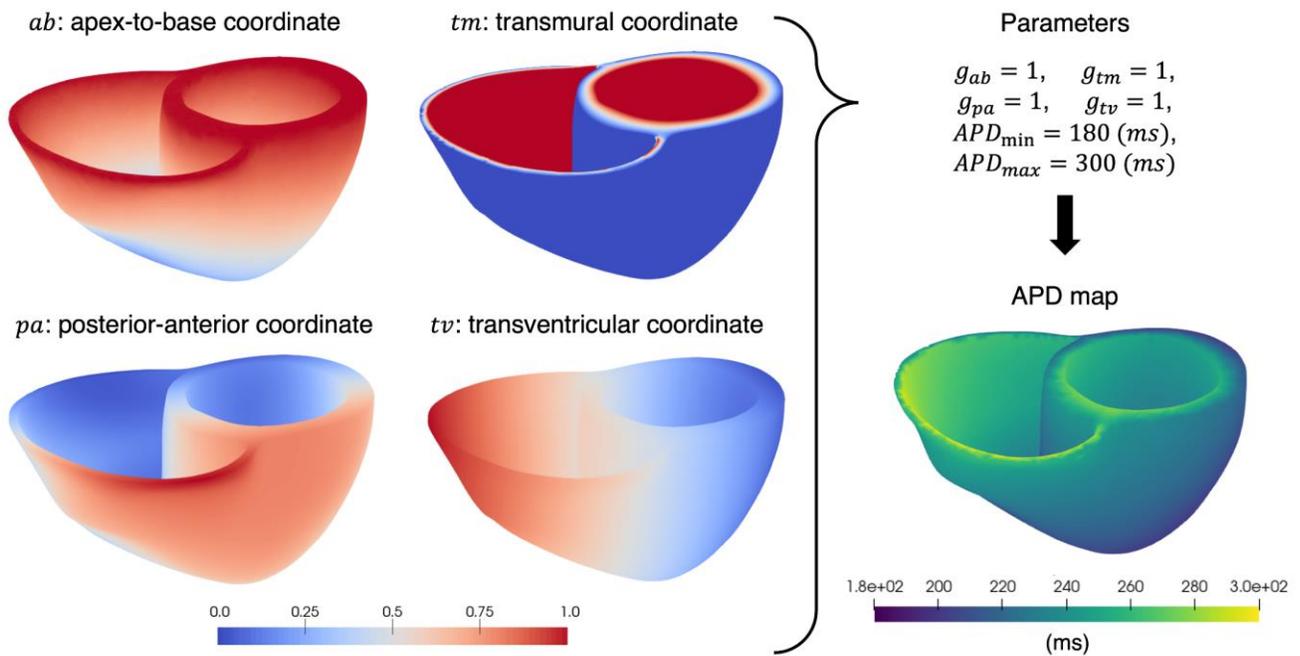

Figure 2. Biventricular coordinates were used to produce a representative APD map, which is generated using a linear combination of gradient weights ($g_{ab}$, $g_{tm}$, $g_{pa}$, and $g_{tv}$) along these coordinates with a specified APD range.

## 2.3 Experimental bases for spatial heterogeneity in action potential duration and ionic conductance

The experimental evidence in predominantly human data shows APD spatial heterogeneities in both magnitude and direction of increase in the transmural, apex-to-base, transventricular, and posterior-to-anterior directions (Table 1). Therefore, assuming significant individual variations in a healthy human population is not unreasonable. Consequently, we allowed the spatial APD gradients to vary in direction and magnitude along these directions in the ventricles during the inference process.

Spatial variations in various ionic currents, underpinning APD heterogeneity, have been documented in the literature (Table 2) in both canine and human data. The most robust evidence for spatial heterogeneity in the transmural, apex-to-base, and transventricular directions is for the slow delayed rectifier potassium current ($I_{Ks}$) and the transient outward potassium current ($I_{to}$) (Table 2). This study considers only modulation through $I_{Ks}$, as it has a larger effect on the APD than $I_{to}$. Considering a single ionic current has the added benefit of precluding spatial variations in other morphological measurements of the action potential apart from the APD (i.e., APD90), such as APD50 and plateau amplitude, which helps to simplify the interpretation of the effects of APD on the T wave morphology in the analysis.

Table 1: Spatial variations in action potential duration (APD), activation recovery intervals (ARIs), and repolarisation time (RT), recorded in the literature in human and canine and showing significant variation in the magnitude and direction of the spatial variation in each of the transmural, apex-to-base, transventricular, and posterior-to-anterior ventricular axes.

| Ventricular axes | Species | Preparation | APD or ARI | RT |
|---|---|---|---|---|
| Transmural | Human (n=10, nine male, one female) (Franz et al., 1987) | In vivo monophasic action potential | - | endo > epi |
| | Human (n=1 normal donor heart) (Opthof et al., 2017) | Langendorff perfusion, monophasic action potential measurements | endo > mid > epi (275 vs 265 vs 255) (LV, normal T wave)<br>endo > epi (315 vs 188) (septal base LV)<br>epi > endo (315 vs 188) (septal mid)<br>endo > epi (315 vs 188) (septal apex)<br>endo > epi (315 vs 188) (freewall, mid, and apex) | No differences. |
| | Human (n=1) Brugada syndrome (Coronel et al., 2005) | | epi > endo > subepi (280 vs 260 vs 240 ms) | endo > epi (220 ms vs 205 ms) |

| | | | Endo > Epi (247 ms vs 242 ms) | Epi > Endo (265 ms vs 247 ms) |
|---|---|---|---|---|
| | Human (n=1) (Conrath et al., 2004) | | | |
| Apex-to-base | Canine (n=7 myocytes) (Szentadrassy et al., 2005) | Ex vivo voltage-clamp measurements | Shorter at the apex by ~50 ms, Base > Apex. Plateau amplitude was lower at the apex by ~15 mV. | - |
| | Human (n=10) (Chauhan et al., 2006) | In vivo transvenous electrode catheters. | Base > apex > mid (280 vs 270 vs 260 ms) on endo<br>Apex > base (300 vs 270 ms) on epi | Base > apex (320 ms vs 310 ms) on endo,<br>Apex > base (350 vs 340 ms) on epi |
| | Human (n=10) (Cowan et al., 1988) | In vivo monophasic action potentials | - | Postero-basal was the earliest to repolarise. Implying Apex > Base |
| | Human (n=1) (Opthof et al., 2017) | Langendorff perfusion | Mid > Base = Apex (252 = 252 vs 188) (septal LV)<br>Base > Mid > Apex (315 vs 252 vs 188) (freewall LV) | Trend (not statistically significant): Apex = Mid > Base (LV and RV, normal T wave) |
| | Human (n=7, four males, three females) (Ramanathan et al., 2006) | In vivo non-invasive ECGI | Base > Apex (anterior, 240 ms vs 230 ms)<br>Base > Apex (posterior, 290 ms vs 270 ms) | Base > apex (anterior, 280 vs 260 ms) |

| | | | | Base > apex (posterior, 260 vs 330 ms) |
|---|---|---|---|---|
| Transventricular | Human (n=10, nine male, one female) (Franz et al., 1987) | In vivo monophasic action potential | Apico-septal > postero-lateral on endocardium, implying Septum > LV, suggesting a gradient in the direction RV > LV | Trend (no statistical significance): Apico-septal and diaphragmatic (RV lateral) are larger than the rest. |
| | Human (n=1 normal donor heart) (Opthof et al., 2017) | Langendorff perfusion | LV basal-anterior > Central-posterior RV Implying LV > RV (298 at basal-anterior LV, 200 ms at central-posterior RV). | No differences. |
| | Human (n=10 with upright T waves) (Cowan et al., 1988) | In vivo monophasic action potentials | - | Anterior and posterior septum is latest to repolarise. Implying Septum > LV/RV |
| | Human (n=7, four males, three females) (Ramanathan et al., 2006) | In vivo non-invasive ECGI | LV > RV (280 vs 250 ms) | LV > RV (340 vs 270 ms) |

| | Human (n=15, 11 males, four females) (Bueno-Orovio et al., 2012) | In vivo invasive electrograms using two catheters | LV > RV (208 vs 197 ms) | |
|---|---|---|---|---|
| Posterior-to-anterior | Human (n=1 normal donor heart) (Opthof et al., 2017) | Langendorff perfusion | Anterior > Posterior (315 vs 188) (basal, LV) | Anterior > Posterior = Lateral (LV only) (315 vs 188 vs 188) |
| | Human (n=7, four males, three females) (Ramanathan et al., 2006) | In vivo non-invasive ECGI | Posterior > anterior (280 vs 240 ms) | Posterior > anterior (350 vs 270 ms) |
| | Human (n=10 with upright T waves) (Cowan et al., 1988) | In vivo monophasic action potentials | - | Postero-basal was earliest to repolarise. Implying anterior > posterior |

Table 2: Ventricular spatial heterogeneity in ionic current magnitude and ionic channel subunit expression from experimental recordings in both human and canine. These data were collected using combinations of three techniques: voltage-clamp, western blotting, and reverse transcription polymerase chain reaction (RT-PCR).

| Ionic currents | Species | Preparation | Channel expression | Peak current and tail current magnitude |
|---|---|---|---|---|

| $I_{to}$ | Human data (n=6, healthy hearts) (Näbauer et al., 1996) | voltage-clamp | Epi > Endo | |
|---|---|---|---|---|
| | Human expression data (n=7), canine functional recordings (n=7 myocytes) (Szentadrassy et al., 2005) | Western blotting in human, voltage-clamp for canine | Kv1.4: 40% base/apex ratio (apex > base), KChIP2: 75% base/apex ratio (apex > base) Kv4.3: No base/apex difference | Canine: peak current apex > base (29.6 +- 5.7 pA/pF vs 16.5 +- 4.4 pA/pF) |
| | Human expression data (n=5), canine functional recordings (n=6 myocytes) (Szabó et al., 2005) | Western blotting in human, voltage-clamp for canine | Epi > Mid | |
| | Human RT-PCR data (n=15) (Gaborit et al., 2007) | RT-PCR data, confirmed by Western blotting | Epi > Endo (KChIP2) | |
| | Human RT-PCR data (n=7, non-failing hearts) (Soltysinska et al., 2009) | RT-PCR data confirmed by Western blotting | Epi > Mid > Endo (KChIP2) | |
| | Human (n=1 normal donor heart) (Opthof et al., 2017) | Langendorff perfusion | LV > RV (Kv1.4) Epi > Endo (KChIP2) | - |

| $I_{Ks}$ | Human expression data (n=7), canine functional recordings (n=7 myocytes) (Szentadrassy et al., 2005) | Western blotting in human, voltage-clamp for canine | KvLQT1: 40% base/apex ratio (apex > base)<br><br>MinK: 70% base/apex ratio (apex > base) | Canine: Both peak and tail currents apex > base (5.61 vs 2.14 pA/pF, 1.65 vs 0.85 pA/pV) |
|---|---|---|---|---|
| | Human expression data (n=5), canine functional recordings (n=6 myocytes) (Szabó et al., 2005) | Western blotting in human, voltage-clamp for canine, comparing EPI and MID regions | Epi > Mid | |
| | Human RT-PCR data (n=7, non-failing hearts) (Soltysinska et al., 2009) | RT-PCR data | Mid > Endo (KCNE1)<br>Mid > Epi (KCNE1) | |
| | Human (n=1 normal donor heart) (Opthof et al., 2017) | Langendorff perfusion | Septum > RV > LV (KvLQT1)<br>Septum > LV > RV (MinK) | - |
| $I_{Kr}$ | Human expression data (n=7), canine functional recordings (n=7 myocytes) (Szentadrassy et al., 2005) | Western blotting in human, voltage-clamp for canine | hERG: No base/apex difference,<br>MiRP1: No base/apex difference | Canine: tail current no base to apex difference |
| | Human expression data (n=5), canine functional recordings (n=6 myocytes) (Szabó et al., 2005) | Western blotting in human, voltage-clamp for canine | There was no significant difference between Epi and Mid | |
| | Human (n=1 normal donor heart) (Opthof et al., 2017) | Langendorff perfusion | hERG: LV > RV <br> MiRP1: No difference in any direction | - |
| $I_{K1}$ | Human expression data (n=7), canine functional recordings (n=7 myocytes) (Szentadrassy et al., 2005) | Western blotting in human, voltage-clamp for canine | Kir2.1: No base/apex difference | Canine: peak current above 40 mV membrane potential apex > base |
| | Human expression data (n=5), canine functional recordings (n=6 myocytes) (Szabó et al., 2005) | Western blotting in human, voltage-clamp for canine | No significant difference between Epi and Mid | |
| | Human RT-PCR data (n=7, non-failing hearts) (Soltysinska et al., 2009) | RT-PCR data | Kir2.1 (KCNJ2): Mid > Endo <br> Kir2.1 (KCNJ2): Mid > Epi | |
| $I_{CaL}$ | Human expression data (n=7), canine functional recordings (n=7 myocytes) (Szentadrassy et al., 2005) | Western blotting in human, voltage-clamp for canine | Alpha_1C: No base/apex difference T | Canine: peak current no base to apex difference |
| | Human RT-PCR data (n=15) (Gaborit et al., 2007) | RT-PCR | Epi > Endo (Cav1.2) | |
| | Human RT-PCR data (n=7, non-failing hearts) (Soltysinska et al., 2009) | RT-PCR data | Cav1.2: Mid > Endo<br>Cav1.2: Mid > Epi | |
| $I_{NaCa}$ | Human RT-PCR data (n=7, non-failing hearts) (Soltysinska et al., 2009) | RT-PCR data | Mid > Endo (NCX1)<br>Mid > Epi (NCX1) | |
| late INa | Human RT-PCR data (n=15) (Gaborit et al., 2007) | RT-PCR | Endo > Epi (Nav1.5) | |
| | Human RT-PCR data (n=7, non-failing hearts) (Soltysinska et al., 2009) | RT-PCR data, confirmed by western blotting | Nav1.5 (SCN5A): Mid > Epi<br>Nav1.5 (SCN5A): Endo > Epi | |
| $I_{NaK}$ | Human RT-PCR data (n=7, non-failing hearts) (Soltysinska et al., 2009) | RT-PCR data | no statistical difference in any direction | |
| $J_{rel}$ | Human RT-PCR data (n=7, non-failing hearts) (Soltysinska et al., 2009) | RT-PCR data | Mid > Endo (RyR2)<br>Mid > Epi (RyR2) | |
| $J_{up}$ | Human RT-PCR data (n=15) (Gaborit et al., 2007) | RT-PCR data | Epi > Endo (SERCA2) | |
| | Human RT-PCR data (n=7, non-failing hearts) (Soltysinska et al., 2009) | RT-PCR data | no statistical difference in any direction | |
| CMDN | Human RT-PCR data (n=15) (Gaborit et al., 2007) | RT-PCR | Epi > Endo (CALM3) | |
| $I_{KATP}$ | Human RT-PCR data (n=7, non-failing hearts) (Soltysinska et al., 2009) | RT-PCR data | Mid > Endo (Kir6.2) Mid > Endo (SUR1) Mid > Epi (SUR1) | |

## 2.4 Reaction-Eikonal model of human cardiac electrophysiology

One of the main hurdles for generating cardiac digital twins is the high computational cost of simulating mechanistic models of cardiac electrophysiology that enable virtual therapy evaluation. To overcome this challenge, we present a novel reaction-Eikonal model paired with a strategy to translate reaction-Eikonal simulations to monodomain simulations. Our reaction-Eikonal model of human cardiac electrophysiology is formulated as follows:

$$U(\boldsymbol{x}, t) = U_{rest} + H\big(t - t_a(\boldsymbol{x})\big)K_{A,\tau}$$
$$A = [APD], \qquad \tau = \min([t - t_a(\boldsymbol{x})], t_{max})$$

*(2)*

where $U(\boldsymbol{x}, t)$ is the membrane potential transient field, which varies spatially over mesh nodes $\boldsymbol{x}$ and temporally over $t$, $U_{rest}$ is the resting membrane potential before the action potential's upstroke, $H$ is the Heaviside function, and square brackets indicate rounding to integer values. The time of electrical activation, $t_a(\boldsymbol{x})$, is calculated by solving the Eikonal model using

$$\begin{cases} \sqrt{\nabla t_a^T(\boldsymbol{x}) \boldsymbol{V} \nabla t_a(\boldsymbol{x})} = 1 & in \ \Omega_{\boldsymbol{x}} \\ t_a(\boldsymbol{x}) = t_i & for \ \boldsymbol{x} = \boldsymbol{y}_i, where \ i = 1..n \end{cases}$$

*(3)*

where $\boldsymbol{V}$ is the conduction velocity tensor (prescribing orthotropic conduction in the fibre, sheet, and normal directions), and $t_i$ is the activation times of the $n$ earliest activation root nodes located at $\boldsymbol{y}_i$. In Equation 2, $K$ is a precomputed lookup table that uniquely maps integer values of APD ($A = [APD]$) and time within the course of the action potential ($\tau$) to a corresponding membrane potential $U$:

$$K : A, \tau \rightarrow U$$

*(4)*

The action potentials $U(\tau)$ were computed by solving the ordinary differential equation.

$$\frac{dU}{d\tau} = I_{stim} + I_{ion}$$

*(5)*

where $I_{ion}$ is the sum of ionic currents in a human-based model of ventricular cardiomyocyte electrophysiology (ToR-ORd) (Tomek et al., 2019) and $I_{stim}$ is a stimulus current. The construction of $K$ will be described in more detail later.

### 2.4.1 Human-based electrophysiology model

Our reaction-Eikonal model makes use of the ToR-ORd (Tomek et al., 2019) model for human cardiomyocyte electrophysiology, which describes ionic currents and calcium signalling in the human cardiomyocyte using a series of Hodgkin-Huxley-type equations and Markov models. The ToR-ORd model was selected for its extensive validation using experimental data for disease phenotypes and drug response (Zhou et al., 2022).

### 2.4.2 Electrotonic coupling effect in the reaction-Eikonal model

To enable the reaction-Eikonal model to mimic the electrotonic coupling effects during the activation sequence in monodomain simulations, we select a stimulus current $I_{stim}$ for the cellular simulations that is equal to an averaged diffusive current $I_{diff}$ ($I_{stim} \equiv I_{diff}$) that was extracted from a monodomain simulation of the activation phase, such that the cellular equation becomes:

$$\frac{dU}{d\tau} = I_{diff} + I_{ion}$$

*(6)*

This use of the diffusive current $I_{diff}$ has been described by Gassa et al. (2021). Conceptually, $I_{diff}$ is a spatially averaged approximation of the diffusion term $\nabla \cdot (\boldsymbol{D}\nabla U)$ in the monodomain equation during the activation sequence (Neic et al., 2017),

$$\chi C_m \frac{\partial U}{\partial t} - \nabla_X \cdot (\boldsymbol{D} \nabla_X U) + \chi I_{ion}(U, \boldsymbol{w}, \boldsymbol{c}) = \chi I_{app}(\boldsymbol{X}, t), \qquad \Omega_0 \times (0, T],$$

$$I_{diff} = \nabla_X \cdot (\boldsymbol{D} \nabla_X U),$$

*(7)*

where $\chi$ is the cardiomyocyte surface-to-volume ratio, $C_m$ is the membrane capacitance per unit area of the average cardiomyocyte, $U$ is the membrane potential, $I_{ion}$ are the sum of ionic currents from the ToR-ORd cell model where $\boldsymbol{w}$ and $\boldsymbol{c}$ are the ionic channel gating variables and intracellular ionic concentrations of the model, $I_{app}$ is the stimulus current applied at 3D ventricular scale, and $\boldsymbol{D}$ is the orthotropic tensor of local conductivities in the reference configuration defined as,

$$\boldsymbol{D} = \sigma_f \boldsymbol{f} \otimes \boldsymbol{f} + \sigma_s \boldsymbol{s} \otimes \boldsymbol{s} + \sigma_n \boldsymbol{n} \otimes \boldsymbol{n},$$

*(8)*

where $\sigma_f, \sigma_s, \sigma_n$ are the conductivities in the fibre $\boldsymbol{f}$, sheet $\boldsymbol{s}$, and normal $\boldsymbol{n}$ directions, respectively. The diffusive currents $I_{diff}$ were extracted per mesh node using the first 100 ms of a monodomain simulation that had been pre-calibrated to the subject's QRS segment (Camps et al., 2023). The extracted currents are then aligned using the upstroke of the local action potential and averaged. We then parameterised a bi-exponential equation, as proposed by Gassa et al. (2021), to this averaged diffusion current signal so that it can be applied as a stimulus for the cellular simulations:

$$I_{diff} = A_1 e^{-\frac{(t-\mu_1)^2}{2\sigma_1^2}} - A_2 e^{-\frac{(t-\mu_2)^2}{2\sigma_2^2}}$$

*(9)*

The amplitude of the parameterised diffusive current equation was then scaled to the minimum value that could fully excite all cell models in the lookup table computation and was used to generate $K$.

In addition to the diffusive current in the activation sequence, the membrane potential field $U(\boldsymbol{x}, t)$ produced by the reaction-Eikonal solution is smoothed spatially to mimic the effect of electrotonic coupling during repolarisation. At each point in time, smoothing was applied for each node in the mesh $(n)$ using a weighted average of the membrane potential of all its adjacent nodes $(m)$, as well as of itself. Let $M$ be the number of adjacent nodes. The smoothed membrane potential $(\overline{U}_n)$ at each node is, then, the weighted sum of membrane potentials from all adjacent nodes $(U_m)$ as well as from itself $(U_n)$,

$$\overline{U}_n = \sum_{m=1}^{M} \left( \frac{k_m U_m}{\sum_{m=1}^{M} k_m + k_n} \right) + \frac{k_n U_n}{\sum_{m=1}^{M} k_m + k_n}$$

*(10)*

where an adjacency weighting factor $(k_m)$ was designed such that the edge vector most aligned with the direction of the highest conduction speed has the highest weighting. Let the edge vector between a node and its adjacent node be $\boldsymbol{p}$, with $\|\boldsymbol{p}\|$ the distance to each adjacent node. Formally, the adjacency weighting factor $(k_m)$ is defined as follows:

$$k_m = \frac{1}{\sqrt{\left( \frac{\max(V_f, V_s, V_n)}{V_f} \boldsymbol{p} \cdot \boldsymbol{f} \right)^2 + \left( \frac{\max(V_f, V_s, V_n)}{V_s} \boldsymbol{p} \cdot \boldsymbol{s} \right)^2 + \left( \frac{\max(V_f, V_s, V_n)}{V_n} \boldsymbol{p} \cdot \boldsymbol{n} \right)^2}}$$

*(11)*

where $V_f, V_s, V_n$ are the conduction speeds along the fibre $\boldsymbol{f}$, sheet $\boldsymbol{s}$, and normal $\boldsymbol{n}$ directions, respectively. Note that $k_m$ is bound by

$$\frac{\min(V_f, V_s, V_n)}{\max(V_f, V_s, V_n)} \frac{1}{\|\boldsymbol{p}\|} \leq k_m \leq \frac{1}{\|\boldsymbol{p}\|}$$

*(12)*

The conduction speeds $V_f, V_s, V_n$ have been previously inferred from the QRS segment of the ECG (Camps et al., 2023).

The membrane potential from the current node is weighted by a self-weighting factor ($k_n$) that was set to a large value ($k_n = 20\ cm^{-1}$), equivalent to a distance to self of 50 micron, bearing in mind that the average element edge length for the reaction-Eikonal simulation is 1500 micron.

The amount of smoothing applied is controlled by the number of repetitions of the operation so that as the number of repetitions increases, the smoothing incorporates more remote effects. To mimic the effect of diffusion, we increased the number of smoothing repetitions linearly throughout the cardiac cycle from one at $t = 0\ ms$ to 40 repetitions by $t = 400\ ms$, at which point the number of repetitions is kept constant at this peak value. The peak repetition number was calibrated to allow the best translation between reaction-Eikonal and monodomain simulations.

### 2.4.3   Lookup table-aided fast computation of cellular electrophysiology

The lookup table $K$ was precomputed by simulating a population of cell models by uniformly sampling I$_{Ks}$ conductance (G$_{Ks}$) varying between 1/50 to 50-fold its baseline value. In order to achieve a sufficiently large range of cellular APD through varying G$_{Ks}$ alone, the baseline ToR-ORd model was modified by scaling the conductance of the rapidly activating delayed rectifier potassium current (G$_{Kr}$) by 70%, the conductance of the slowly activating delayed rectifier potassium current (G$_{Ks}$) by 5-fold (Doste et al., 2022), and decreasing the time constant of L-type calcium channel activation ($\tau_{jca}$) from 75 to 60 ms (Tomek et al., 2020). Each cell model was simulated at a cycle length of 800 ms for 100 beats stimulated by the diffusive current ($I_{diff}$), and the resulting population was calibrated by discarding any models inconsistent with biomarker ranges for the action potential and calcium transient (Appendix Table A1) (Passini et al., 2019). The APD range of the resulting lookup table was [180, 300] ms. The shape of the lookup table $K$ was $N \times T$, where $N = 120$ for each integer APD value and $T = 600$ for each time point in the action potential transient up to 600 ms of simulation.

## 2.5   Electrocardiogram simulation

Standard 12-lead ECGs were simulated from the reaction-Eikonal membrane potential simulations $U(\boldsymbol{x}, t)$ using the pseudo-ECG method (Gima & Rudy, 2002; Camps et al., 2021). The simulated ECGs are normalised with respect to the R-progression of the clinical data (Camps, Berg et al., 2023) to ensure comparability. Body surface potentials, $\Phi$, were calculated at the electrode locations ($\boldsymbol{x}'$) using:

$$\Phi(\boldsymbol{x}') = \sum_{j=1}^{\mathrm{N}_{src}} -\boldsymbol{D}_j (\nabla U)_j \left[ \nabla \frac{b_j}{r_j} \right],$$

*(13)*

where $(\nabla U)_j$ is the spatial gradient of the membrane potential over the $j$ th tetrahedral element, $\boldsymbol{D}_j$ is the diffusivity tensor at the $j$th element, $b_j$ is the normalised volume scaling factor for the $j$ th element, $r_j$ is the Euclidean distance from the centroid of the $j$-th element to the electrode location($\boldsymbol{x}'$), and $\mathrm{N}_{src}$ is the total number of tetrahedral source elements. The pseudo-ECG method provides a fast and simple evaluation of the normalised ECG without major loss of morphological information compared with bidomain simulations (Wallman et al., 2012; Camps et al., 2021). When evaluating the ECG recordings from the reaction-Eikonal simulations, we consider the diffusive term to be the identity matrix, whereas, in the monodomain simulations, the diffusive term is obtained from the orthotropic diffusivities in the geometry. We demonstrate that the effect of this difference in formulation is small enough to enable the later translation between reaction-Eikonal and monodomain simulations (Figure A2).

We analysed the effect of the choice of cellular model on the T wave morphology using a unipolar electrogram approximation, as described by Potse et al. (2009). This consisted of subtracting a simulated action potential by itself time-shifted by 20 ms. We compared the effects of using the ToR-ORd versus the Mitchell-Schaeffer (MS) cellular models (Mitchell & Schaeffer, 2003) on T wave morphology (Figure A3). The MS model is a two-current action potential model that was considered in similar studies (Gillette et al., 2021).

### 2.5.1    Sensitivity analysis of ECG characteristics to APD gradients

Using the reaction-Eikonal model, we performed a global sensitivity analysis of T wave morphology biomarkers to the parameters that prescribe spatial heterogeneity in APD: $g_{ab}$, $g_{tm}$, $g_{pa}$, $g_{tv}$, $APD_{min}$, and $APD_{max}$. Following the evidence regarding spatial APD heterogeneities in the ventricles (Section 2.3), the parameters were sampled in the following ranges: [-1,1], [-1,1], [-1,1], [-1,1], [180-230], [270-300], respectively. The sensitivities were evaluated as Sobol indices (Sobol', 2001), and the sampling was done using the Saltelli sampling sequence (Saltelli, 2002; Saltelli et al., 2010), considering uniform priors for all parameters. The quantities of interest were the mean QT interval, T-peak to T-end interval, T wave amplitude, and T wave polarity across leads I, II, V1, V2, V3, V4, V5, and V6 using gradient-based evaluation methods. Total and first-order sensitivity effects were evaluated.

## 2.6   Inference of activation and repolarisation parameters

### 2.6.1   Inference of conduction speeds and root nodes to match the QRS complex data

The biventricular electrical activation pattern is inferred from the 12-lead QRS ECG segment using the methods found in Camps et al. (2021, 2023), which are summarised hereafter. An Eikonal model is used to simulate the electrical activation in the biventricular mesh. The activation pattern is determined by selecting a series of earliest activation sites (root nodes) from a series of discretely defined locations on the endocardial surfaces, as well as by the conduction velocities defined along the fibre, sheet, and sheet-normal directions. The time at which each root node activates, relative to the moment of His bundle activation, is prescribed using a human-based physiologically-informed Purkinje tree network (Camps et al., 2023). The discrepancy between simulated QRS and target input signals is calculated using the dynamic time warping-based algorithm proposed by Camps et al. (2021). Inference with the sequential Monte Carlo approximate Bayesian computation (SMC-ABC) algorithm with discrete parameter space capabilities developed by Camps et al. (2021) provides an optimised population of models with specific combinations of conduction speeds and root node selections that were best able to reproduce the input 12-lead QRS ECG complexes. The most relevant hyperparameters of the inference algorithm for this study are the sampling rate per iteration, which affects the convergence speed; the discrepancy cut-off, which controls how well the inferred simulations match the data; and the uniqueness threshold, which controls the variability in the inferred population.

The outputs of this inference consist of the best-match activation times $t_a(x)$ across the entire biventricular domain, and the spatially constant conduction velocities $V_f, V_s, V_n$ in the fibre, sheet, and sheet-normal directions, respectively. For this study, we build on the results presented in (Camps et al., 2023) for control subject 2. Thus, the root node and conduction speed inference results were not discussed.

### 2.6.2   Inference of APD gradients to match ST-T ECG signals

As for the activation phase, the SMC-ABC algorithm was implemented to infer a set of parameters that determine ST-T ECG segment signals given the activation sequence in the ventricles. The parameter set consists of the APD ranges and spatial gradients: $g_{ab}, g_{tm}, g_{pa}, g_{tv}, APD_{min}, APD_{max}$.

A starting population (of parameter sets) of size (n=256) was created using Latin Hypercube Sampling of the parameters, and the objective function ($\epsilon$) was evaluated per sample in the population as:

$$\epsilon = 100 \times \frac{1}{L} \sum_{i=0}^{L} (1 - PCC_i)^2 + 2 \times \frac{\frac{1}{L} \sum_i^L RMSE_i}{\max(|ECG_{clinical}|)}$$

*(14)*

where $PCC_i$ is the Pearson correlation coefficient at lead $i$ evaluated between each simulated and clinical lead, then averaged over all leads ($L = 8$), RMSE is the root mean squared error for each lead $i$ between each simulated and clinical lead, then averaged over all leads, and $\max(|ECG_{clinical}|)$ is the maximum amplitude across all leads of the normalised clinical ECG data.  Note that the RMSE is dimensionless because it has been calculated based on normalised ECG signals.

The parameter space for the repolarisation gradients was defined as continuous. However, we know that similar parameter configurations can yield indistinguishable simulation results when evaluated. Thus, to speed up the inference process and promote convergence to different solutions in the final population, we discretised the parameter space. The $APD_{min}$ and $APD_{max}$ were discretised to a resolution of 2 ms, and the gradient parameters ($g_{ab}, g_{pa}, g_{tm}$, and $g_{tv}$) were discretised to a resolution of 0.1 gradient units.

The inference hyperparameters were set to 0.5 for the discrepancy cut-off and 50% for the uniqueness threshold. This meant that the inference process would continue until either all simulated parameter sets in the population had a discrepancy of less than 0.5 or the proportion of the population with unique parameters fell below 50%. The value of the discrepancy cut-off was calibrated through trial and error and was deliberately set to a small value so that the method would reduce the discrepancy as much as possible while preserving computational efficiency, which is ensured by the uniqueness threshold. The values for all hyperparameters for the inference can be found in the code repository.

The variability of the inferred population represents parameter uncertainty given the subject's ECG and biventricular anatomy. This uncertainty can be propagated to drug simulations by translating from reaction-Eikonal to monodomain simulations.

## 2.7   Translation from Reaction-Eikonal to Monodomain simulations

In addition to the diffusive and smoothing strategies, as previously described (Section 2.4.2), we calibrated the orthotropic diffusivity parameters $\sigma_f$, $\sigma_s$, $\sigma_n$ in the monodomain model to achieve desired conduction velocities using a tissue slab simulation (Niederer et al., 2011). The diffusivity parameters were optimised using a bounded gradient-descent method from the SciPy library,  to achieve fibre and sheet-normal conduction velocities of 65 cm/s and 48 cm/s from Taggart et al. (2000), and to achieve sheet conduction velocity of 44 cm/s, as inferred from the QRS complex. Conduction velocities were evaluated using first-order finite difference evaluations of the activation time map in the middle of the tissue slab.

We extracted the $G_{Ks}$ scaling factors that correspond to the fitted APD maps of the inferred population and embedded them in the monodomain simulations. Closest-point interpolation was used to translate from the coarser mesh (1500 micron edge length) used for reaction-Eikonal simulations to the finer mesh (900 micron edge length) used for monodomain simulations.

## 2.8   Evaluation of digital twins using virtual drug simulations

We simulated the effect of Dofetilide, an $I_{kr}$ blocker, at seven doses and compared its QT prolongation effect against clinical data (Vicente et al., 2015). For this, we randomly selected 5% of models (n=13) from the inferred population as the baseline. Dofetilide effect was simulated for the following concentrations: 0.05 nM, 1 nM, 2 nM, 3 nM, 4 nM, 5 nM, and 6 nM, which corresponded to reducing the $I_{Kr}$ current by 40%, 50%, 60%, 66%, 70%, 73%, 75% (Crumb et al., 2016).

## 2.9   Computation and Software

This study extends our previous methodologies (Camps et al., 2021, 2023) for the generation of cardiac digital twins from CMR and 12-lead ECG. The code for all these previous methodologies has been re-structured in modules with a unified coding style (Python/Numpy) together with the new developments proposed in this study to form a complete cardiac digital twining pipeline that utilises the QRS and T wave from 12-lead ECG data. This new structure was designed to ease software repurposing and extension by others and is further described in Appendix A.2. The code is self-contained and can be found at github.com/juliacamps/Cardiac-Digital-Twin with a working example and tutorial (made available after publication).

The inference process required approximately 15 hours of computation time (Appendix A.6, Fig. A5). The inference was run on a machine with 2 CPUs, each with an AMD EPYC 7313 16-core processor. These results (Appendix A.6) illustrate an example of how to configure the tool's hyperparameters depending on the requirements of each use case.

The Python SciPy library was used for the optimisation process to translate the conduction speeds to monodomain conductivities. Sampling and Sobol indices calculations were done using the SALib Python library (Herman & Usher, 2017; Iwanaga et al., 2022). Monodomain simulations were performed using the Alya finite element solver (Santiago et al., 2018) using 512 cores of a high-performance computer (JURECA), which required 30 minutes to simulate 800 ms.

# 3   Results

## 3.1   Inference results using the reaction-Eikonal models

The inferred population was able to match both QRS and T wave morphologies in the clinical data (Figure 3B), achieving a Pearson correlation coefficient of 0.93 ± 0.0003 and root mean squared error (RMSE) of 0.097 ± 0.0008. This RMSE value was considered small, considering that the maximum amplitude of the QRS complex is

~1.5. The simulated ECGs from the final population showed the presence of a bifid, or notched, T wave, which was likely due to a notched first derivative of phase three of the ToR-ORd cell model, which is reported in more detail in Appendix A.5.

The inferred population had a final discrepancy of 0.74. The inference process was terminated by the uniqueness threshold rather than by the discrepancy cut-off, meaning that the final population had less than 50% uniqueness in the parameter sets while the discrepancy was still above the desired cut-off of 0.5. This was a good outcome since it meant that the discrepancy cut-off of 0.5 was well-chosen to push the inference process to match the clinical data as well as possible while preserving the uniqueness of parameter sets in the population.

The inference process explored a wide range of T wave biomarker values over 50 iterations (grayscale gradients in Figure 3) before arriving at the final population. The mean and standard deviation of the inferred parameters in the final population were $APD_{min} = 216 \pm 0\,ms$, $APD_{max} = 298 \pm 4.8\,ms$, $g_{ab} = 0.86 \pm 0.12$, $g_{pa} = -0.79 \pm 0.14$, $g_{tv} = 0.22 \pm 0.09$, $g_{tm} = 0.21 \pm 0.14$. Interestingly, the T-peak to T-end (Tpe) biomarker converged into two different clusters of values, while all other biomarkers seemed to have a single cluster of values in the final population.

**A**

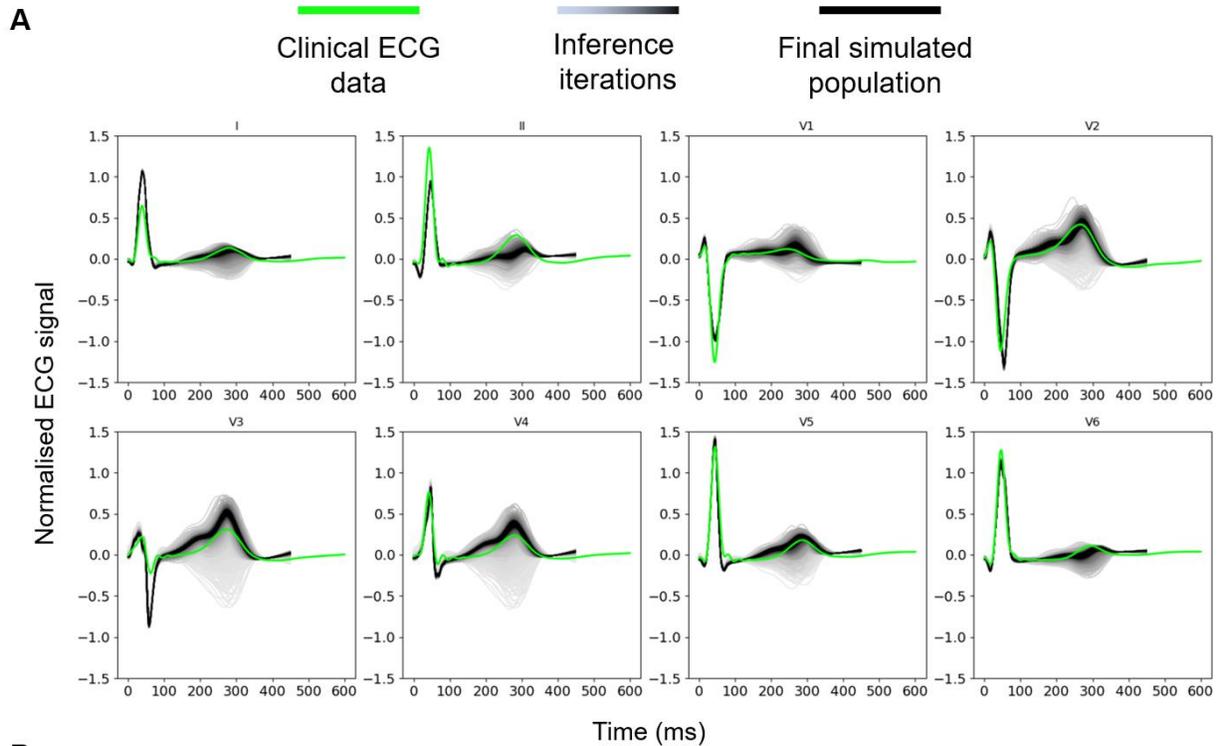

**B**

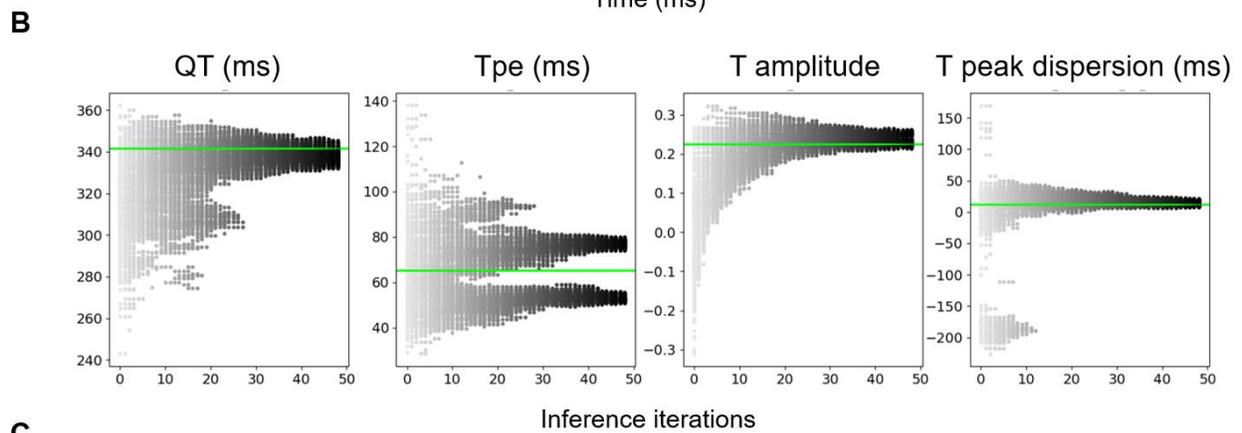

**C**

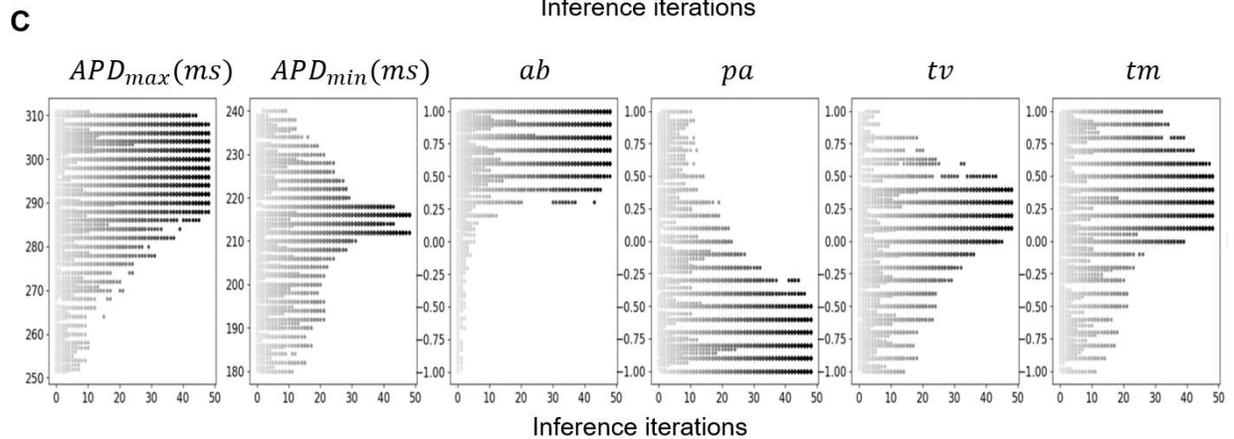

Figure 3: Inference iterations effectively explore T wave biomarker space. Clinical subject ECGs are shown in green, successive inference iterations are shown in increasing blackness, and the final inferred population is shown in black. A) The range of T wave morphologies explored by the inference process, converging on the population with the best match to clinical data. B) QT interval, T-peak to T-end interval (Tpe), average T wave amplitude, and dispersion of T peak timing between leads V3 and V5 converging over successive inference iterations (grey to black) to match clinical values (marked in green horizontal line). C) Progression of the parameter space over successive inference iterations. This parameter space was composed of $APD_{\max}$ (maximum action potential duration), $APD_{min}$ (minimum action potential duration), $g_{ab}$ (APD gradient in the apex-to-base direction), $g_{pa}$ (APD gradient in the posterior-to-anterior direction), $g_{tv}$ (APD gradient in the transventricular direction) and $g_{tm}$ (APD gradient in the transmural direction). Hyperparameter values: sampling rate = 64 samples/iteration, and desired cut-off = 0.5, uniqueness threshold: 50%.

The model with the lowest discrepancy value was selected from the inferred population as the best-match model for further analyses. The best-match parameter values were: $APD_{min} = 216\ (ms)$, $APD_{max} = 294\ (ms)$, $g_{ab} = 1, g_{pa} = -1, g_{tv} = 0.3, g_{tm} = 0.1$. In the best-match model (Figure 4A), the APD map showed a significant posterior-to-anterior gradient, with larger APD on the posterior than the anterior, a transventricular gradient with larger APD in the right than in the left ventricle, and an apex-to-base gradient with larger APD at the base than at the apex. The effect of the transmural gradient was less prominent with larger APD on the endocardium than the epicardium. These gradients were inverted in the G_{Ks} scaling factor map, reflecting the inverse relationship between I_{Ks} magnitude and APD in the cellular models.

## 3.2  Translation from reaction-Eikonal to monodomain simulations

The best-matched reaction-Eikonal model was translated to monodomain simulations by prescribing the spatial gradient of G_{Ks} scaling factors in the monodomain simulation. The translated monodomain simulation showed a good agreement with reaction-Eikonal simulated ECG and clinical ECG in terms of T wave morphology and polarity (Figure 4B). Activation and repolarisation maps were also well-matched in pattern between the reaction-Eikonal and monodomain simulations (Figure 4C).

**A**

APD from model with best match to ECG (ms)   GKs scaling factor from model with best match to ECG

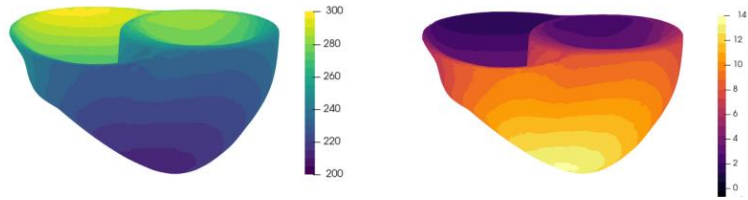

**B**

Clinical ECG     Best reaction-Eikonal ECG simulation     Translated monodomain ECG simulation

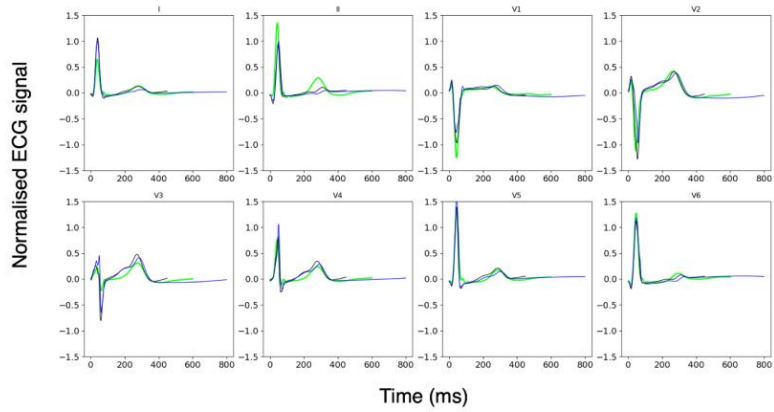

**C**

Reaction-Eikonal Simulation     Monodomain Simulation

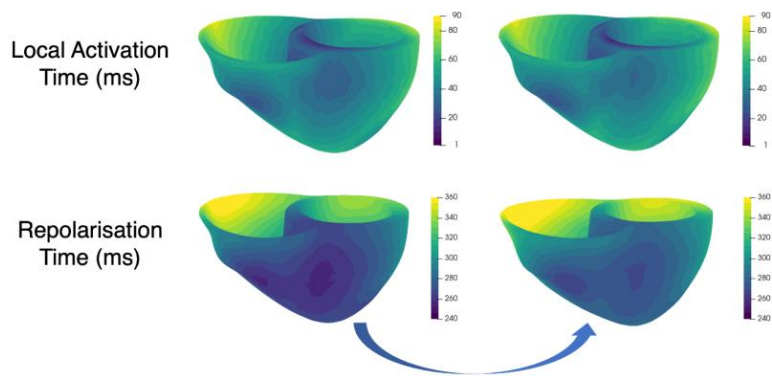

Figure 4: (A) Inferred map of APD with corresponding scaling factor field for the G$_{Ks}$ parameter. The G$_{Ks}$ scaling factor map was used to translate the inferred model to monodomain simulation. (B) Reaction-Eikonal match to clinical ECG vs Monodomain match to clinical ECG after translation. (C) Simulated local activation and repolarisation time maps for the reaction-Eikonal (left) and monodomain (right) models.

Translation between reaction-Eikonal and monodomain simulations was a key step in the realisation of virtual drug evaluations using cardiac digital twins. To match the conduction velocity characteristics of reaction-Eikonal and monodomain simulations, diffusivities of the monodomain simulation ($\sigma_f$, $\sigma_s$, $\sigma_n$) were calibrated to 1.263, 0.966, and 0.967 cm$^2$/s. The diffusive currents that were extracted (Figure 5A - grey traces) were predominantly biphasic (98.77%), with positive monophasic (1.2%) and negative monophasic (0.03%) curves in the minority. The parameters of the bi-exponential function (Equation 9) were fitted to the averaged diffusive current (Figure 5A - black trace) to produce the fitted function (Figure 5A - green trace). The fitted parameters were A1=13.8 µA/µF, A2=14.4 µA/µF, $\mu_1$=14.1 ms, $\mu_2$=15.3 ms, $\sigma_1$=1.9 ms, and $\sigma_2$=1.8 ms. The fitted diffusive current function was scaled to an amplitude of 11 µA/µF, which was the minimum amplitude that could elicit action potentials for the entire calibrated cell population of models (Figure 5B).

**Diffusive current extraction and cellular lookup table generation**

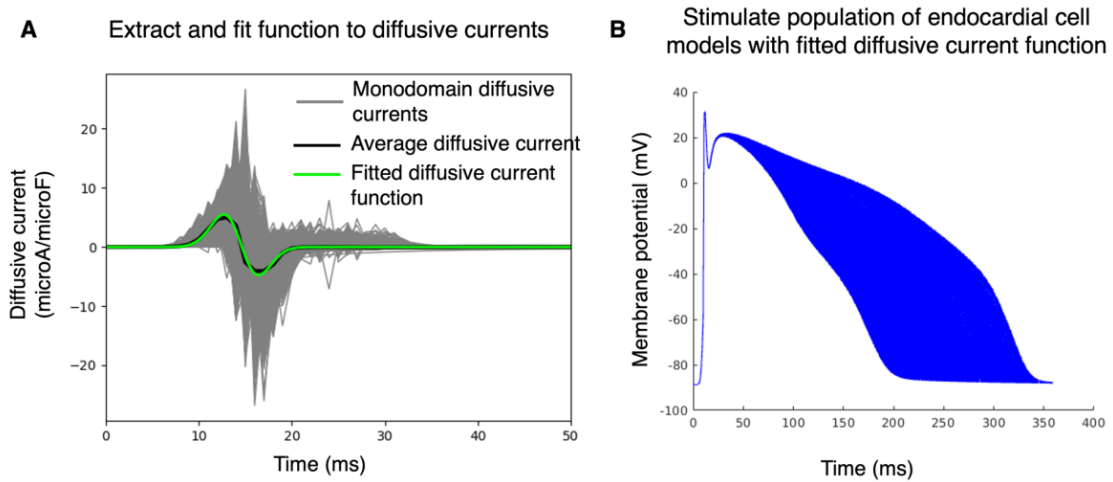

Figure 5: Diffusive current extraction and cellular lookup table generation. (A) Diffusive currents extracted from monodomain simulations (grey) were averaged (black) and fitted to a bi-exponential curve (green). (B) The fitted diffusive current function was then used to stimulate a population of endocardial cell models that are calibrated to experimental data ranges for action potential and calcium transient biomarkers. This population was used to create the lookup table for APD values.

Spatio-temporal smoothing of the reaction-Eikonal simulations was tuned to forty rounds of smoothing (Section 2.10) to allow a close match in T wave morphology and amplitude between the reaction-Eikonal simulations (Figure 6D, x40 black trace) and the monodomain simulations (Figure 6D blue trace). The necessity of using dynamic smoothing (linearly increasing the number of rounds of smoothing over the cardiac cycle) and the inclusion of orthotropic smoothing based on local fibre, sheet, and normal orientations is illustrated in Figure 6. Dynamic smoothing (Figure 6C) and orthotropic smoothing (Figure 6D) were added sequentially to demonstrate their effects (Figure 6B). In summary, dynamic smoothing of the reaction-Eikonal simulation allows a better match to monodomain T wave amplitude, while orthotropic smoothing preserves the transmural pattern of repolarisation even at high smoothing counts.

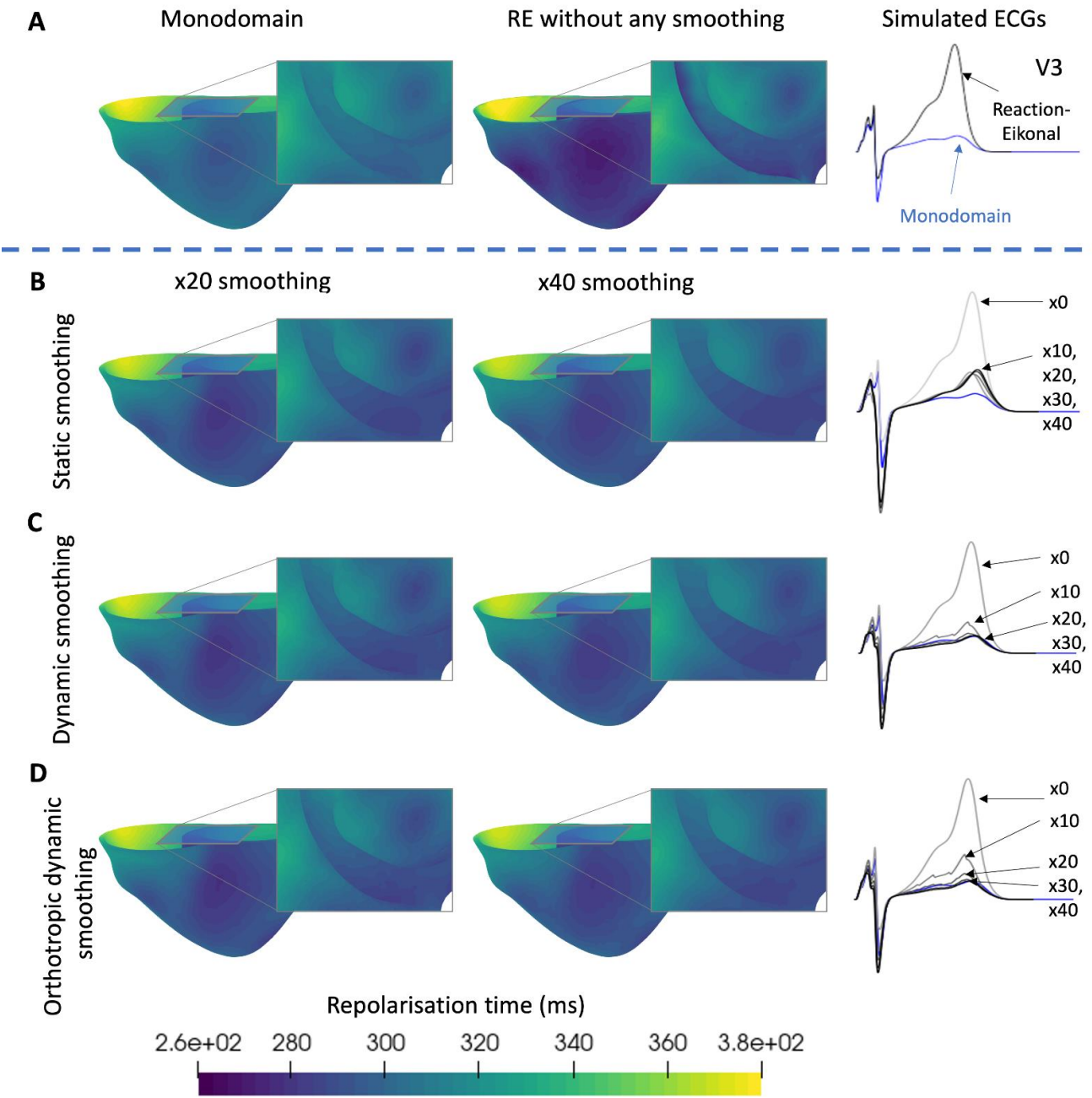

Figure 6: Effect of different smoothing strategies on repolarisation pattern and ECG. (A) Without any smoothing, the repolarisation pattern of the reaction-Eikonal (RE) simulation (middle) occurs earlier and has a larger

transmural gradient of repolarisation times (basal view insets) compared with monodomain, and this is reflected by a much higher simulated T wave amplitude in the RE (black trace) compared with monodomain (blue trace). (B) When isotropic smoothing is applied uniformly throughout the cardiac cycle (static smoothing) to the RE simulation, as the number of rounds increases, the amplitude of the simulated T wave decreases but converges to an amplitude that is larger than the monodomain simulation. In addition, as the smoothing rounds increase, the transmural pattern of repolarisation becomes homogeneous (x40 basal view inset), which deviates from the pattern observed in the monodomain simulations. (C) When smoothing is increased linearly over time (dynamic smoothing), the simulated T wave morphology of the RE (black traces) becomes much more similar to the monodomain simulation (blue trace). However, the transmural repolarisation pattern suffers from the same deviation as in the static smoothing case (x40, basal view inset). (D) Finally, with the inclusion of orthotropic smoothing, the transmural repolarisation pattern of the RE mimics that of monodomain simulation even at high smoothing rounds (x40 basal view inset), while the RE simulated T wave morphology (black traces) continues to match well with the monodomain simulation (blue trace).

## 3.3   Virtual drug evaluations using a cardiac digital twin

A 5% subset of the reaction-Eikonal models in the final inferred population (n=13) was randomly selected for simulations of the effects of the $I_{Kr}$ blocker Dofetilide (Figure 7). These reaction-Eikonal models were translated into monodomain simulations to create a baseline population of ventricular electrophysiological models (black traces Figure 7A and C). The simulated effect of Dofetilide application showed a graded increase in mean QT interval (Figure 7B, blue dots) and T-peak to T-end duration (Figure 7B, black dots), which matched the expected trend seen in clinical data (Johannesen et al., 2014; Vicente et al., 2015; Crumb et al., 2016) (Figure 7B, yellow and cyan traces, respectively).  The steeper increase in T-peak to T-end duration in the clinical data compared with our simulations suggests spatially heterogeneous effects of the drug's action that have not been accounted for in our modelling.

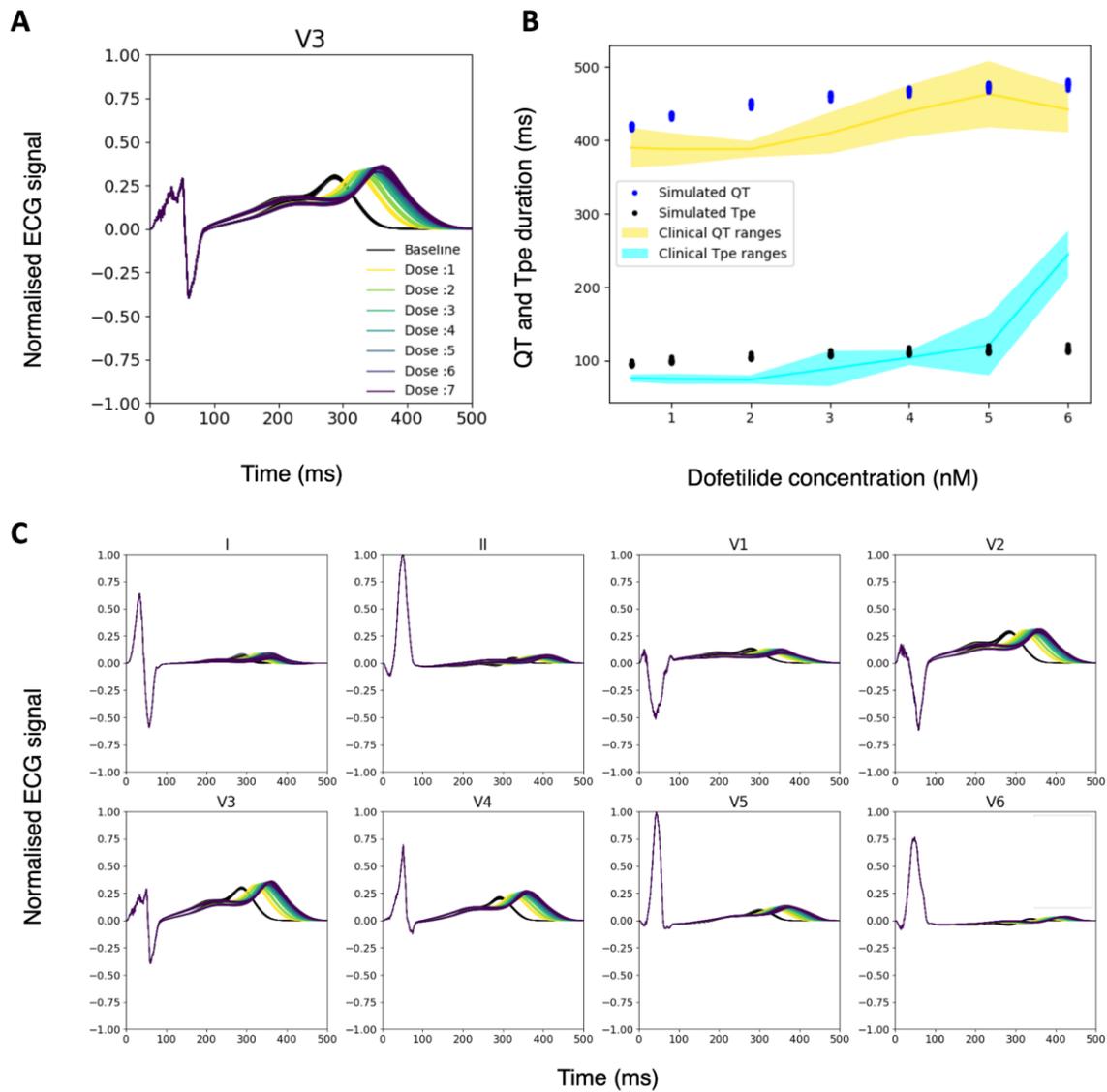

Figure 7: Dofetilide virtual evaluation using the cardiac digital twin. (A) Graded QT-prolonging effect of seven doses of the $I_{Kr}$ blocker Dofetilide on the selected inferred models (n=13), dose 1: 0.05 nM, dose 2: 1 nM, dose 3: 2 nM, dose 4: 3 nM, dose 5: 4 nM, dose 6: 5 nM, dose 7: 6 nM (B) showing good match to trends seen in clinical data (Vicente et al., 2015) on the dose-response of QTc and T-peak to T-end durations, showing contours of standard deviations of the biomarkers. Averaged clinical ranges for QTc were in the range of 390±27 at the lowest dose to 442±31 ms at the highest, and for T-peak to T-end from 76±5 to 245±32 ms. (C) Full 8-lead simulated ECGs showing a consistent QT-prolongation effect of Dofetilide across all leads.

## 3.4 Global sensitivity analysis of T wave biomarkers to APD parameters using reaction-Eikonal models

Results of sensitivity analysis (Figure 8) performed using the reaction-Eikonal model showed strong positive correlations of QT with maximum APD, T wave amplitude and polarity with the apex-to-base and transmural gradients. Global sensitivity analysis showed that maximum APD is by far the most important determinant of the QT interval, while the transmural and apex-to-base gradients were most important for T wave amplitude and polarity. T-peak to T-end duration was sensitive to all gradient parameters, with the transmural gradient having the largest effect by a small margin. Our analysis highlights the non-specificity of the T-peak to T-end duration in the ECG to repolarisation heterogeneity in any particular ventricular spatial axis.

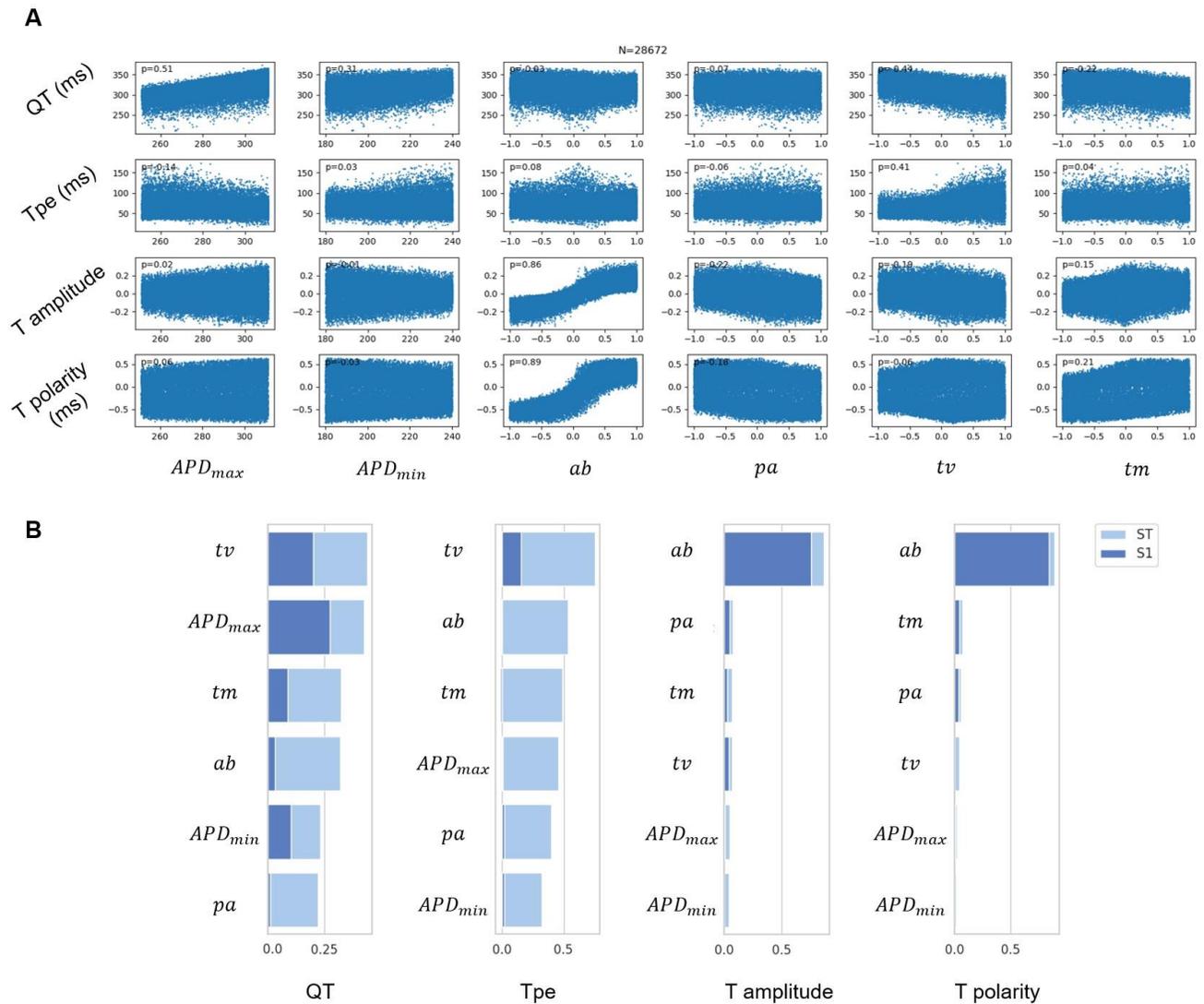

Figure 8: Sensitivity of T wave biomarkers to APD spatial heterogeneities. (A) Scatter plot with Pearson correlations of QT interval, T-peak to T-end (Tpe) interval, T wave amplitude, T wave polarity, and dispersion of T peak timing between leads V3 and V5 to maximum APD ($APD_{max}$), minimum APD ($APD_{min}$), apex-to-base gradient ($ab$), posterior-to-anterior gradient ($pa$), transventricular gradient ($tv$), and transmural gradient parameters ($tm$). (B) Ranked total Sobol index (ST) and first-order effects index (S1) from global sensitivity analysis.

# 4   Discussion

We present a novel pipeline for virtual therapy evaluation with uncertainty propagation using cardiac digital twins informed by clinical CMR and ECG, building on previous work on inferring activation characteristics from 12-lead ECG. Key novelties of this pipeline are i) personalised repolarisation heterogeneities, which were enabled by ii) a reaction-Eikonal model designed to capture human ventricular electrophysiology and iii) mimic electrotonic coupling to allow translation to monodomain simulations, for iv) personalised *in silico* drug evaluations with uncertainty propagation. We have demonstrated this here using Dofetilide at seven doses. This pipeline paves the way for large-scale application to clinical databases, such as those found in the UK Biobank and evaluating a larger suite of drug compounds.

While calibration of repolarisation characteristics using the 12-lead ECG has been achieved in a previous study (Gillette et al., 2021), their ventricular model relied on phenomenological models of cellular electrophysiology. The lack of biophysical ionic detail and human relevance presents a significant hurdle for such methods to be used for drug therapy testing and disease mechanism explorations (Passini et al., 2017; Zhou & Wang et al., 2022). In this study, we provide a novel strategy where we incorporate a state-of-the-art human-based ventricular electrophysiological model.

Our inference method showed a good match to the ST-T ECG segment in the final inferred population. Rather than selecting a single best match, the SMC-ABC method allows us to recover a population that has similarly good matches to clinical data (Figure 3). This uncertainty was then propagated to Dofetilide evaluations, where we showed qualitative agreement with the literature (Crumb et al., 2016) (Figure 5). In this study, we demonstrated this capability using a population size of 13. Nevertheless, this can be easily scaled up according to the specifications of each use case.

The inference process explored a range of T wave biomarker space before converging to the final population (Figure 3). The termination criteria of the inference algorithm controlled the match between the final inferred population and the clinical ECG data. Depending on the assumptions of each use case, these criteria could be relaxed to terminate the inference at an earlier iteration with a larger variability of ECGs in the final population (a 'greyer' region in Figure 3) in acknowledgement of measurement uncertainties in ECG data acquisition. The discrepancy metric can also be designed to overcome limitations in modelling. For example, in this study, the use of the pseudo-ECG method only provides meaningful ECG amplitudes that are relative between leads, which prompted the design of the discrepancy metric to penalise morphological differences and R wave progression errors while ignoring absolute differences in amplitude through normalisation of both the simulated and the clinical ECGs (Camps et al., 2023).

Our reaction-Eikonal model also allows global sensitivity analysis of the effect of APD gradient parameters on the ST-T ECG segment morphology (Figure 6). These results confirm that it is possible to anticipate the parameter ranges for the inference from the clinical ECG characteristics. For example, the maximum APD parameter will define the range of QT intervals that the inference will be able to explore. Meanwhile, the apex-to-base APD gradients will influence the amplitude and polarity of the T wave the most, suggesting that knowing the polarity of the T wave beforehand can allow us to constrain the parameter space accordingly.

A key technical hurdle in enabling large-scale digital twins based in silico trials is enabling fast digital twin generation while preserving necessary biophysical details for therapy intervention. In this paper, we demonstrate an effective pairing of a fast reduced-order model (reaction-Eikonal) with a reaction-diffusion model to achieve this goal. The faithful and efficient translation between the two is of crucial importance for the integrity of the pipeline. This is achieved by a two-fold development in our reaction-Eikonal model: 1) using human-based cellular electrophysiology to generate APD look-up tables, and 2) the implementation of a diffusive current stimulus and a spatio-temporal smoothing for the reaction-Eikonal model (Figure 7). The incorporation of these strategies proved critical for the successful translation between reaction-Eikonal and monodomain 12-lead ECG simulations (Figure 4).

The scalability and computational efficiency of the method allow for the creation of large healthy virtual cohorts with real patient-specific ventricular electrophysiological function for *in silico* clinical trial evaluations of drug effects. Our pipeline could generate a population of cardiac digital twins per subject in 46 hours, with the following timeline breakdown: 5.5 hours for 3D anatomical reconstruction (Banerjee et al., 2021), 24 hours for Bayesian inference of the activation properties (Camps et al., 2022), 15 hours for inference of repolarisation properties, and 1.5 hours for the calibration of the monodomain conductivities and parameterisation of the stimulus diffusive current. The *in silico* evaluations of drug effect requires 0.5 hours per therapy configuration. With readily available supercomputing resources, this process could be parallelised per subject to tackle as large a dataset as resources permit, using paired CMR and ECG data from large datasets such as the UK Biobank, of which there are data in the tens of thousands (Sudlow et al., 2015; Littlejohns et al., 2020). Furthermore, the computational cost of the inference process can be modulated using the sampling rate and the termination criteria hyperparameters (Appendix A.6), which allows tailoring the approach to computational resource availability.

In a similar context of exploiting clinical data to inform clinical decision-making, machine learning tools can also represent the functional and anatomical variability in the human population (Beetz et al., 2022). These machine learning strategies, however, are constrained by the available data such that only predictions within the bounds of previously observed behaviour can be trusted. On the other hand, mechanistic modelling and simulation, such as those presented here, can evaluate never-before-seen therapeutic options and produce new data for informing machine learning models and for informing clinical decision-making (Boyle et al., 2019; Margara et al., 2022; Monaci et al., 2023).

## 4.1  Limitations

The presented pipeline, as well as similar studies (Gillette et al., 2021), assumes that the activation and repolarisation sequences are sufficiently independent that the inference of their properties can be made in sequence. This assumption reduces the parameter space considerably, thus reducing the computational cost and complexity of the inverse problem while still enabling the finding of plausible solutions with a good match to the clinical data (Figure 3). However, given sufficient resources, inferring both activation and repolarisation characteristics simultaneously could widen the functional variability of the inferred population. An evaluation of the effect of this on improving therapy predictions should be investigated.

The ECG simulations in this study are based on approximate electrode locations and use a pseudo-ECG formulation. These choices could potentially introduce some errors in the simulated ECGs (Multerer & Pezzuto, 2021; Nagel et al., 2022). The use of the pseudo-ECG method has been shown to be sufficient for providing realistic morphology of the T wave while having limitations on the amplitude when compared with bidomain simulations (Wallman et al., 2012; Camps et al., 2021), and such limitation is mitigated in this study through normalisation of the pseudo-ECG signal as well as the raw clinical ECG data. In addition, the modularity of the code structure allows other ECG calculation methods, such as the boundary element method, to be easily inserted into the pipeline. The uncertainty of electrode positioning and its effect on digital twinning and drug predictions can be incorporated into the parameter uncertainty by introducing an additional location error sampling step before the ECG simulations take place. These investigations could further improve the personalisation and uncertainty quantification aspects of the work in the future.

While in this paper, we allow only the $I_{Ks}$ conductance to vary spatially, there is evidence of other ionic involvement in the spatial repolarisation heterogeneity (Table 2). However, the majority of the experimental evidence points towards heterogeneity in all four directions in the $I_{Ks}$ and $I_{to}$ currents. In our cellular simulations, the $I_{Ks}$ current had a much larger effect on the APD than $I_{to}$, thus prompting our decision to vary only $I_{Ks}$. In addition, since the $I_{to}$ current has a significant effect on the morphology of phase one of the action potential and, therefore, on the QRS morphology, any future explorations of spatial variation in $I_{to}$ would need to infer both activation and repolarisation characteristics simultaneously. Other morphological aspects of the action potential, such as APD50 and the plateau amplitude, and its influence on T wave morphology should also be investigated in future improvements, especially if multiple current heterogeneities were to be introduced.

A notched (bifid) T wave morphology was consistently observed in the simulated ECG population, where a positive deflection appears in the ST-T segment before the main T wave peak. These notches were not present in another similar study that used the Mitchell-Schaeffer (MS) description of the cellular model (Gillette et al., 2021). This prompted an investigation where we repeated the reaction-Eikonal simulations with the MS model instead of the ToR-ORd, and the bifid T wave morphology disappeared (Figure A.4). From our preliminary analysis, it seemed that his phenomenon resulted from a similar notch in the first derivative of the ToR-ORd action potential, which is not existent in the smoother MS model (Figure A.4). Given that the baseline ToR-ORd model has been calibrated using a multi-objective genetic algorithm, and extensively validated against experimental data, we propose that this limitation could be overcome by including T wave morphology smoothness as an additional objective in an updated calibration of the baseline model. Finally, while the final population showed a good match to the clinical ECG, our findings could be limited by the assumption of linear changes in APD in the four ventricular coordinates. However, until experimental data becomes available at higher spatial resolutions, it remains difficult to justify and parameterise the presence of higher-order and/or non-monotonic gradients.

## 5  Conclusions

We present a novel pipeline for generating cardiac digital twins for virtual therapy evaluation. This pipeline generates biophysically detailed cardiac digital twins from available CMR and ECG recordings with representations of electrophysiological uncertainty. The digital twins were capable of reproducing subject-specific ECG phenotypes and reproducing clinically expected behaviour when simulating the effects of Dofetilide at different doses. Our pipeline automatically propagates uncertainty in the final population to *in silico* drug evaluations to inform any decisions based on the simulated outcomes. The development of translation capabilities between reaction-Eikonal and monodomain simulations decreases computational cost while enabling drug safety and efficacy evaluation capabilities, thus enabling scalable digital twinning technology for future expansion and application to *in silico* clinical trials.

## 6  Conflicts of Interest

The authors declare that the research was conducted without any commercial or financial relationships that could be construed as a potential conflict of interest.

## 7  Author Contributions

**JC:** Conceptualization, Methodology, Software, Investigation, Formal analysis, Validation, Visualization, Writing - original draft, Writing - review & editing. **ZJW:** Conceptualization, Methodology, Software, Investigation, Formal analysis, Validation, Visualization, Writing - original draft, Writing - review & editing. **RD:** Software, Formal analysis, Writing - review & editing. **MH:** Methodology, Formal analysis, Writing - review & editing. **BL:** Methodology, Formal analysis, Writing - review & editing. **JT:** Methodology, Formal analysis, Writing - review & editing. **KB:** Formal analysis, Writing - review & editing. **AB-O:** Conceptualization, Methodology, Formal analysis, Writing - review & editing. **BR:** Conceptualization, Resources, Writing - review & editing, Supervision.

## 8  Acknowledgements

This work was funded by an Engineering and Physical Sciences Research Council doctoral award, a Wellcome Trust Fellowship in Basic Biomedical Sciences to Blanca Rodriguez (214290/Z/18/Z), the CompBioMed 2 Centre of Excellence in Computational Biomedicine (European Commission Horizon 2020 research and innovation programme, grant agreement No. 823712), the Australian Research Council Centre of Excellence for Mathematical and Statistical Frontiers (CE140100049), an Australian Research Council Discovery Project (DP200102101), by the Queensland University of Technology (QUT) through the Centre for Data Science. The computation costs were incurred through PRACE ICEI projects (icp013 and icp019), which provided access to Piz Daint at the Swiss National Supercomputing Centre, Switzerland, and the JURECA machine at the Juelich Supercomputing Centre, Germany.

The authors thank Dr Mariano Vázquez and the other members of ELEM and the Barcelona Supercomputing Centre for providing research access to the Alya simulation software used for the monodomain simulations conducted in this study.

# A. Appendix

## A.1 Digital twinning pipeline workflow overview

The digital twinning pipeline combines a subject's CMR and ECG recordings to estimate a set of root nodes, tissue conductivities, and action potential duration gradients (APD) using a reaction-Eikonal simulation with a ToR-ORd action potential model. These inferred parameters are then translated to monodomain equivalents to enable realistic drug therapy evaluations. The workflow is summarised below and illustrated in Figure A1. The pipeline first reconstructs a 3D geometry of the subject's heart (Banerjee et al., 2021). Secondly, it uses Eikonal simulations to infer a set of earliest activation sites (root nodes in the Purkinje system) and conduction velocities that reproduce the subject's clinical QRS complex (Camps et al., 2021). Next, this process is continued by using reaction-Eikonal simulations to infer the repolarisation APD gradients that recover the subject's ST-T ECG signals. Then, a random 5% set of these parameters is translated to monodomain equivalents, and a drug therapy is evaluated using multiscale simulations of its ionic effects on the ToR-ORd cellular models. The inference method of both the activation and repolarisation characteristics follows the strategy described by Camps et al. (2021).

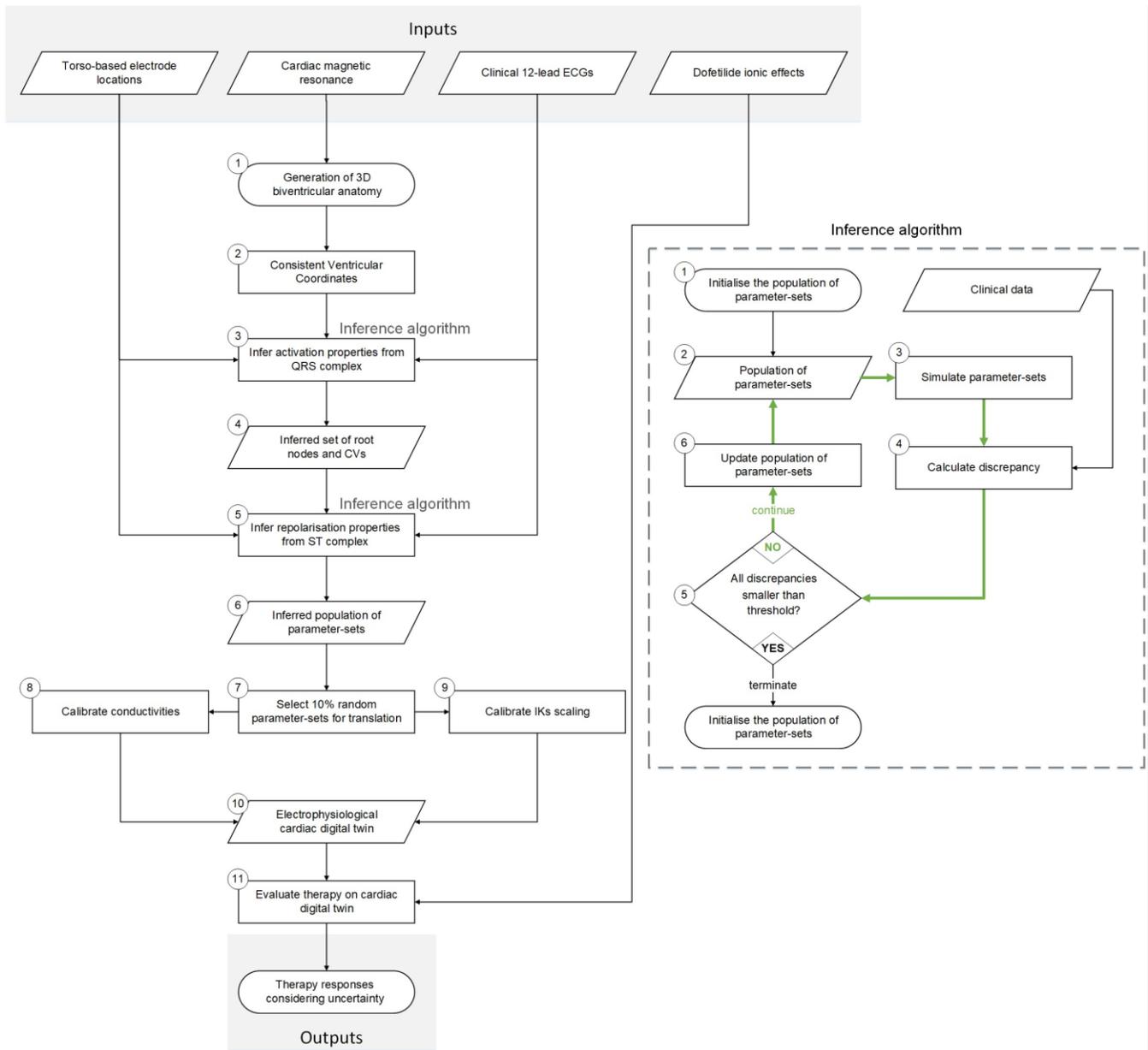

Figure A1. Digital twin generation pipeline flow chart. The flow chart outlines the subprocesses comprising the electrophysiological calibration process for a subject given their CMR data, torso-based electrode locations, and clinical ECG. First, the 3D biventricular geometry is reconstructed from the CMR data. Secondly, the consistent ventricular coordinates are generated from the biventricular geometry. Thirdly, the inference (following the sub-

diagram for the inference algorithm) of the activation properties from the QRS complex retrieves a set of root nodes and conduction velocities. Fourthly, the inference (as in the sub-diagram of the inference algorithm) of repolarisation properties builds on the results of the activation inference process to determine the action potential duration of 90% (APD90 or just APD for the purpose of this study) gradients that reproduce the clinical ST-T ECG signals. These processes produce a population of conduction velocities, root nodes, and $I_{Ks}$ scaling maps, from which 5% are randomly kept for translation to monodomain parameters. The conduction velocities are used to calibrate conductivities, and the $I_{Ks}$ scaling maps are corrected using prior knowledge of the differences between reaction-Eikonal and monodomain simulations. These produce a set of models which all together compose a digital twin from the current subject since they represent the uncertainty in the model's parameters. Finally, the selected therapy (dofetilide) is evaluated in this set of models that compose the digital twin through multiscale simulations of the drug's effects. The arrows determine the sequence of events in the flow chart. The shapes symbolise parallelograms – data; rectangles – processes; rhomboids – decisions. The grey background is used for the input and output data in the pipeline.

## A.2   Code structure of the digital twinning pipeline

The code has been packaged for ease of adoption and future extension. The code is solely in Python (except for monodomain simulations done in Alya), with no dependencies beyond standard Python libraries. All modules are included in the repository of the publication: github.com/juliacamps/Cardiac-Digital-Twin.

The code is divided into modules with distinct responsibilities to facilitate integration with other existing workflows. These modules and their responsibilities are as follows:

- **Geometry:** classes for the geometrical information of the ventricles.
- **Conduction system:** determines how the conduction system should be handled for a given geometry.
- **Cellular:** contains the action potential simulations required by the reaction-Eikonal.
- **Propagation:** contains implementations of different models for the propagation of the electrical wavefront in the ventricles, such as the Eikonal and the reaction-Eikonal models. This does not contain the monodomain implementation, which was simulated using the Alya software package.
- **Electrophysiology:** defines how the cellular and propagation modules interact.

- **Electrocardiogram:** is responsible for calculating and processing the ECG recordings.
- **Simulation:** determines what data will be simulated using the Electrophysiology and Electrocardiogram modules.
- **Discrepancy:** contains different strategies for calculating discrepancy metrics that compare clinical ECG data with simulated ECG data.
- **Evaluation:** determines what should be the output from evaluating a set of parameter values (e.g. discrepancy to clinical data, ECG signal, repolarisation map, etc.) and links the classes of the Simulation and Discrepancy modules.
- **Sampling:** classes for implementations of the inference and sensitivity analysis methods that interface with the Evaluation module.
- **Adapter:** handling class for all the input and sampled parameters. This module controls which parameters are being sampled (*theta*) and makes sure that they are forwarded to the correct modules.

These modules are coupled with each other as follows (**Module name:** main function name):

(Glossary) In the code, *theta* refers to the parameters being sampled, while *parameter* refers to the combination of sampled and prescribed parameters.

- **Sampling:** sample_theta(*population size*)
    - **Evaluation:** evaluate_theta(*theta*)
        - **Discrepancy:** evaluate_metric(*simulation data*)
        - **Adapter:** translate_theta_to_parameter(*theta*)
        - **Simulation:** simulate(*parameter*)
            - **Electrocardiogram:** calculate_ecg(*electrophysiology data*)
                - **Geometry:** *None*
            - **Electrophysiology:** simulate_electrophysiolgy(*parameter*)
                - **Propagation:** simulate_propagation(*parameter*)
                    - Geometry: *None*
                        - **Conduction system:** generate_Purkinje(*ab, rt, tm, tv*)
                        - **Cellular:** generate_action_potential(*APD*)

Generic and shared functions across modules can be found in the utils or input-output (io) modules. The code contains several main scripts for different use cases. Each of them initialises and links classes from these modules explicitly to serve as templates for future use cases. For example, to include a scar and perform inference on its conduction properties, it would be necessary to extend the geometry module to have a class with a scar and to incorporate those parameters in the propagation module and main script.

## A.3   Calibration ranges for the ToR-ORd model

Table A1 provides the calibration ranges obtained from Passini et al. (2019) for the ToR-ORd models in the look-up table (Section 2.4.3) used by the inference method.

Table A1. Biomarker ranges for calibrating the population of ToR-ORd models. APD[x] - action potential duration from peak of action potential to x% of repolarisation. dVdt_max - maximum first derivative of the action potential (upstroke velocity). V_peak – peak membrane potential. RMP – resting membrane potential, evaluated as the minimum membrane potential at steady-state. CTD[x] – calcium transient duration from peak of transient to x% of decay.

| Cellular biomarkers | Calibration ranges |
|---|---|
| APD40 (ms) | [85, 320] |
| APD50 (ms) | [110, 350] |
| APD90 (ms) | [180, 440] |
| Triangulation (APD90-APD40) (ms) | [50, 150] |
| dVdtmax (mV/ms) | [100, 1000] |
| Vpeak (mV) | [10,55] |
| RMP (mV) | [-95, -80] |
| CTD50 (ms) | [120, 420] |
| CTD90 (ms) | [220, 785] |

## A.4   Electrocardiogram calculation differences

As previously stated in Section 2.5, this work employs different formulations for the computation of the pseudo-ECG for the simulations of the reaction-Eikonal vs monodomain models. Here, we show a comparison between calculating the ECG using each of these approaches to justify that for the purposes of the study, the simplified version of the formulation used for reaction-Eikonal simulations can be considered as an approximation of the full pseudo-ECG formulation, which is employed by the monodomain simulations.

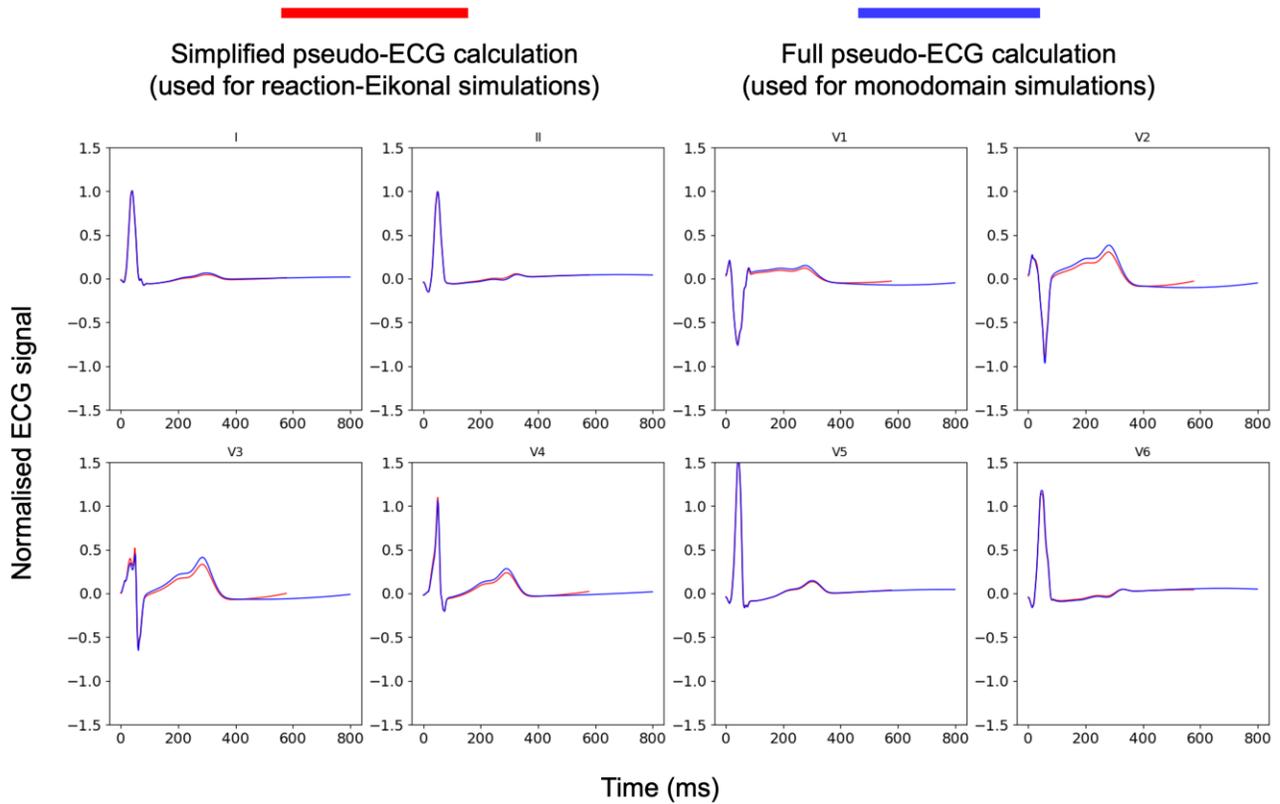

Figure A2: Comparison of pseudo-ECG calculation using an identity tensor for the diffusivity tensor for reaction-Eikonal (red trace), versus using the true diffusivity tensor in monodomain simulation (blue trace). The two traces are very similar, with small differences in T wave amplitude in leads V1 to V3.

## A.5   Investigations of bifid T waves in reaction-Eikonal simulations

We compared the effect of using the ToR-ORd (Tomek et al., 2019) cellular model vs the Mitchell-Schaeffer cellular model (MS) on T wave morphology (Mitchell & Schaeffer, 2003). We calculated the first derivative of the simulated action potentials and saw non-monotonic changes in the gradient of the ToR-ORd model that are significantly less pronounced in the MS model (Figure A3, first column). We then approximated unipolar electrograms for the ToR-ORd and MS cell models by calculating the difference between a baseline action potential and a time-shifted action potential that has been delayed by 20 ms (Figure A3 second column). The ToR-ORd unipolar electrogram showed the bifid morphology while the MS one did not, highlighting the effect of the non-monotonic gradient of the ToR-ORd action potential on the ECG. We also repeated the inference process using either the MS or ToR-ORd models and showed that the simulated ECGs were only notched when using ToR-ORd (Figure A3 third column).

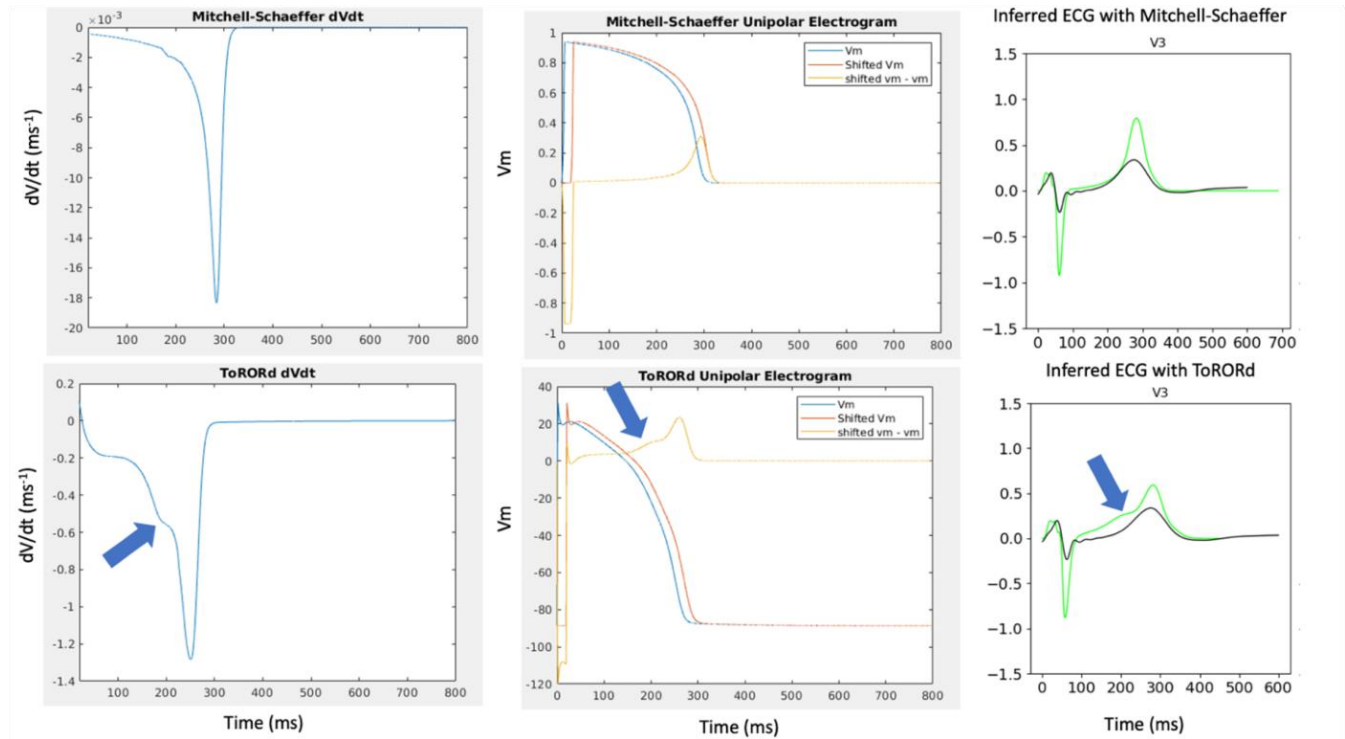

Figure A3: Evaluation of the first derivative over time of the ToR-ORd and Mitchell-Schaeffer (MS) action potentials and saw non-monotonic changes in the gradient of the ToR-ORd model that is significantly less pronounced in the MS model. Furthermore, we approximated unipolar electrograms for the ToR-ORd and MS cell models by calculating the difference between a baseline action potential and a time-shifted action potential that has been delayed by 20 ms. The ToR-ORd unipolar electrogram showed the bifid morphology while the MS one did not, highlighting the effect of the non-monotonic gradient of the ToR-ORd action potential on the ECG

## A.6   Investigations of the computational cost of the inference of repolarisation properties

The inference results reported in the main manuscript (Fig. 3) constitute an exhaustive run of the inference considering a small sampling rate (64 samples per iteration) and a small discrepancy cut-off of 0.5. These results served to explore the limits of the inference machinery. Here, we report an additional two configurations of these hyperparameters to provide a less computationally intensive use case. In the first configuration, we increased the sampling rate from 64 to 96 samples per iteration and the discrepancy cut-off from 0.5 to 1 (Fig. A4).  This change reduced the computational cost of the inference from 37 (Fig.3) to 4 (Fig. A4) hours and reduced the number of iterations of the inference process from 50 (Fig.3) to 9 (Fig. A4). The inference process is sped up by both sampling more samples at a time and setting an easier discrepancy target.

**A**

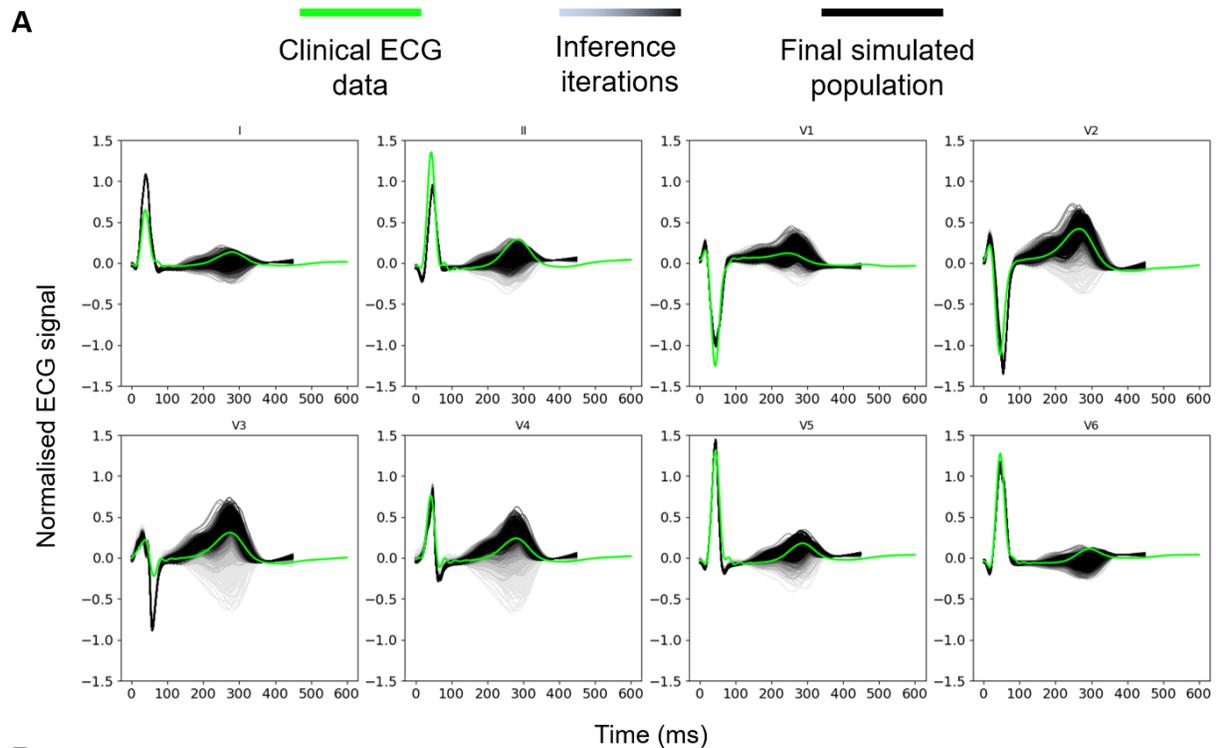

**B**

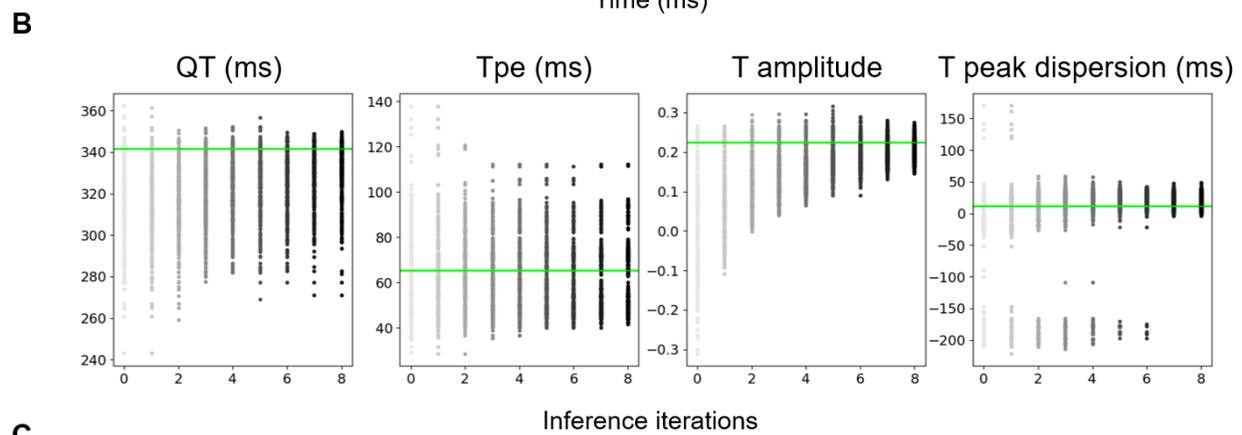

**C**

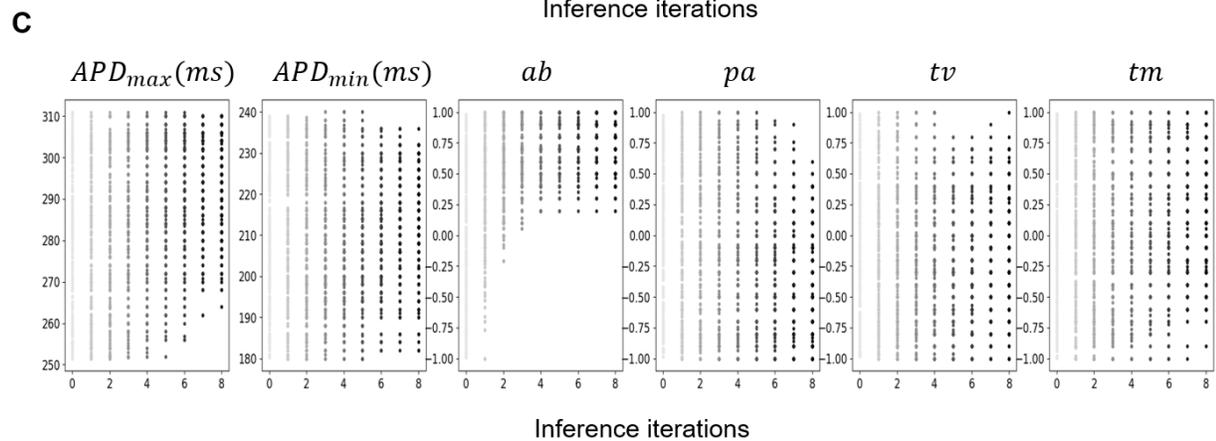

Figure A4: Inference iterations effectively explore T wave biomarker space. Analogous to Figure 3 but with hyperparameter values for a sampling rate of 96 samples/iteration and a desired discrepancy value of 1 (discrepancy-unit). Clinical subject ECGs are shown in green, successive inference iterations are shown in increasing blackness, and the final inferred population is shown in black. A) The range of T wave morphologies explored by the inference process, converging on the population with the best match to clinical data. B) QT interval, T-peak to T-end interval (Tpe), (normalised) average T wave amplitude, and dispersion of T peak timing between leads V3 and V5 converging over successive inference iterations (grey to black) to match clinical values (marked in green horizontal line). C) Progression of the parameter space over successive inference iterations. This parameter space was composed of $APD_{\max}$ (maximum action potential duration), $APD_{min}$ (minimum action potential duration), $g_{ab}$ (APD gradient in the apex-to-base direction), $g_{pa}$ (APD gradient in the posterior-to-anterior direction), $g_{tv}$ (APD gradient in the transventricular direction) and $g_{tm}$ (APD gradient in the transmural direction). Hyperparameter values: sampling rate = 96 samples/iteration, and desired cut-off = 1, uniqueness threshold: 50%.

The second configuration kept the sampling rate at 96 samples/iteration but set the discrepancy cut-off to 0.77, which is larger than the final discrepancy result of the exhaustive run (0.74). (Fig. A5). This was done to estimate how long the inference process would take if the target was similar to the actual goodness of the match achieved by the exhaustive inference process. This inference process was able to reach the desired discrepancy after 15 hours and 20 iterations.

**A**

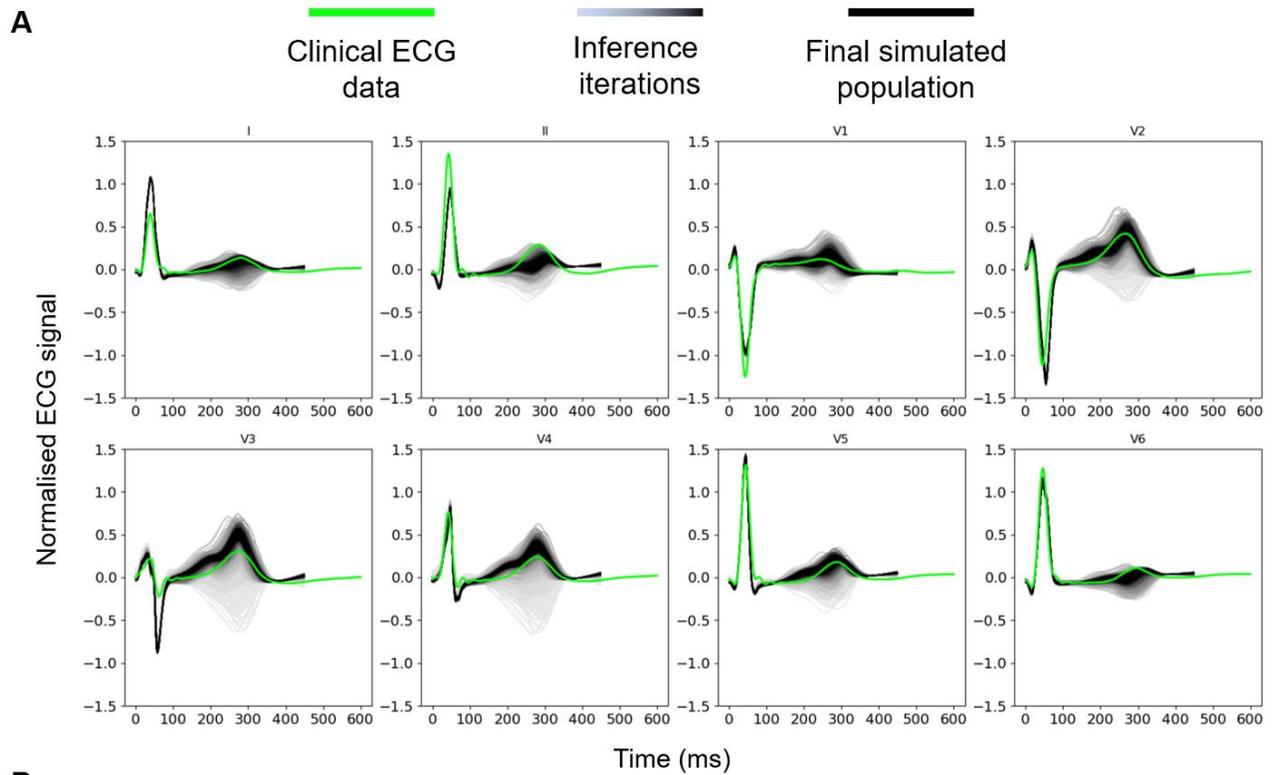

**B**

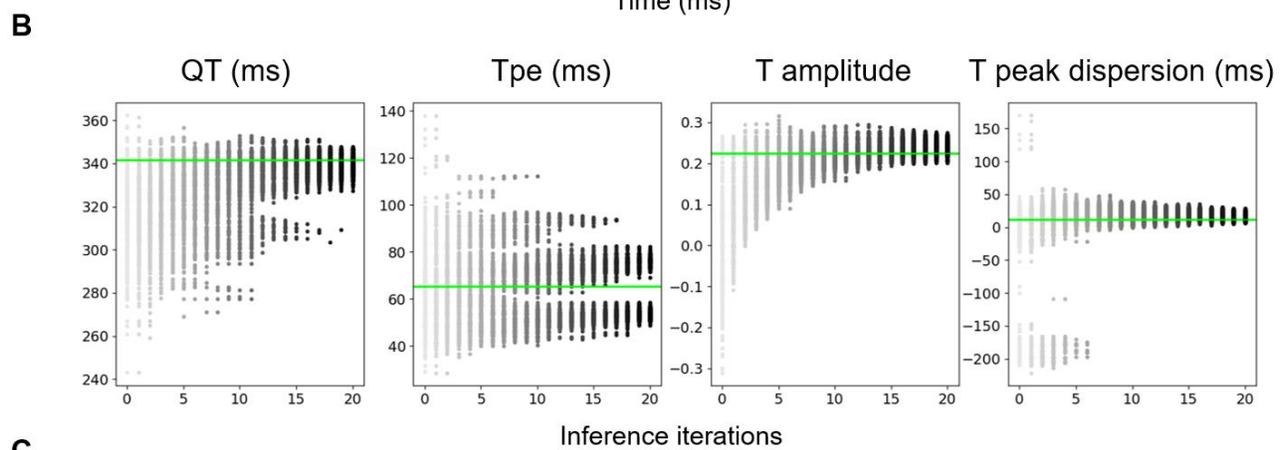

**C**

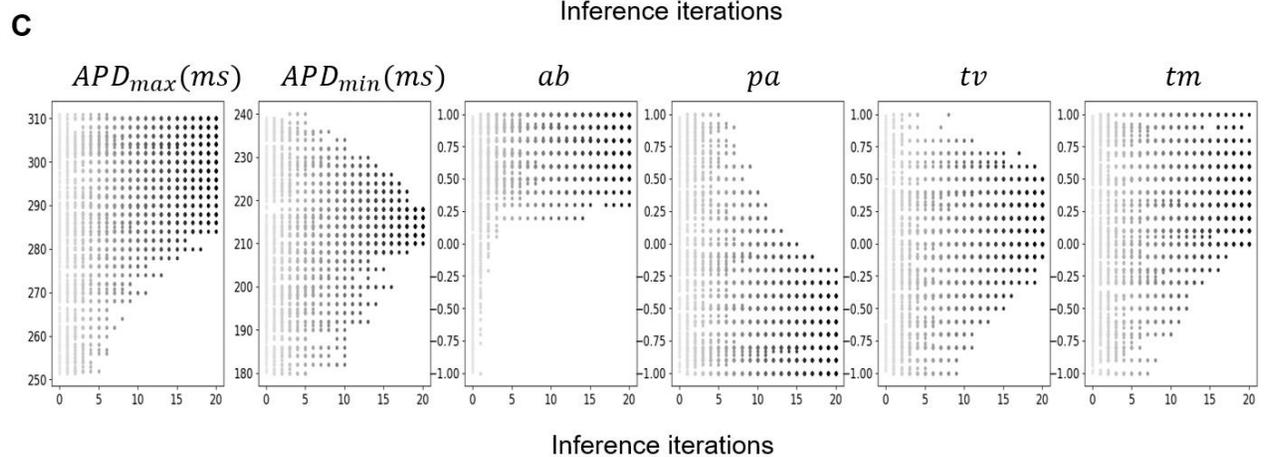

Figure A5: Inference iterations effectively explore T wave biomarker space. Analogous to Figure 3 but with hyperparameter values for a sampling rate of 96 samples/iteration and a desired discrepancy value of 0.77 (discrepancy-units). Clinical subject ECGs are shown in green, successive inference iterations are shown in increasing blackness, and the final inferred population is shown in black. A) The range of T wave morphologies explored by the inference process, converging on the population with the best match to clinical data. B) QT interval, T-peak to T-end interval (Tpe), (normalised) average T wave amplitude, and dispersion of T peak timing between leads V3 and V5 converging over successive inference iterations (grey to black) to match clinical values (marked in green horizontal line). C) Progression of the parameter space over successive inference iterations. This parameter space was composed of $APD_{\max}$ (maximum action potential duration), $APD_{min}$ (minimum action potential duration), $g_{ab}$ (APD gradient in the apex-to-base direction), $g_{pa}$ (APD gradient in the posterior-to-anterior direction), $g_{tv}$ (APD gradient in the transventricular direction) and $g_{tm}$ (APD gradient in the transmural direction). Hyperparameter values: sampling rate = 96 samples/iteration, desired cut-off = 0.77, uniqueness threshold: 50%.